\documentclass[journal,twocolumn,10pt]{IEEEtran}

\usepackage{cite}
\usepackage{amsmath,amssymb,amsfonts}
\usepackage{algorithmic}
\usepackage{algorithm}
\usepackage{graphicx}
\usepackage{textcomp}
\usepackage{xcolor}
\usepackage{subfigure}
\usepackage{graphicx}
\usepackage{booktabs}
\usepackage{multirow}
\usepackage{setspace}
\usepackage{stfloats}

\newtheorem{theorem}{\underline{Theorem}}

\ifCLASSINFOpdf
\else
\fi

\begin{document}
 
\title{\huge Towards Secure ISAC Beamforming:\\ How Many Dedicated Sensing Beams Are Required?}

\author{Fanghao Xia, Zesong Fei,~\IEEEmembership{Senior~Member,~IEEE}, Xinyi Wang,~\IEEEmembership{Member,~IEEE}, Nanchi~Su,~\IEEEmembership{Member,~IEEE}, \\
	Zhaolin Wang,~\IEEEmembership{Member,~IEEE}, Yuanwei Liu,~\IEEEmembership{Fellow,~IEEE}, and Jie Xu,~\IEEEmembership{Fellow,~IEEE}
	
	
	\thanks{Fanghao Xia, Zesong Fei, and Xinyi Wang are with the School of Information and Electronics, Beijing Institute of Technology, Beijing 100081, China (e-mail: xiafanghaoxfh@163.com, feizesong@bit.edu.cn, bit\_wangxy@163.com.).}

	\thanks{Nanchi Su is with the Guangdong Provincial Key Laboratory of Aerospace Communication and Networking Technology, Harbin Institute of Technology (Shenzhen), Shenzhen 518055, China (e-mail: sunanchi@hit.edu.cn).}

	\thanks{Zhaolin Wang and Yuanwei Liu are with the Department of Electrical and Electronic Engineering, The University of Hong Kong, Hong Kong (e-mail: zhaolin.wang@hku.hk, yuanwei@hku.hk).}

	\thanks{Jie Xu is with the School of Science and Engineering (SSE), Shenzhen Future Network of Intelligence Institute (FNiiShenzhen), and Guangdong 		Provincial Key Laboratory of Future Networks of Intelligence, The Chinese University of Hong Kong (Shenzhen), Guangdong 518172, China (E-mail: 		xujie@cuhk.edu.cn).}

%
%
%

%

}

\maketitle

\begin{abstract}
\textcolor{black}{In this paper, sensing-assisted secure communication in a multi-user multi-eavesdropper integrated sensing and communication (ISAC) system is investigated. Confidential communication signals and dedicated sensing signals are jointly transmitted by a base station (BS) to simultaneously serve users and sense aerial eavesdroppers (AEs). A sum rate maximization problem is formulated under AEs' Signal-to-Interference-plus-Noise Ratio (SINR) and sensing Signal-to-Clutter-plus-Noise Ratio (SCNR) constraints. A fractional-programming-based alternating optimization algorithm is developed to solve this problem for fully digital arrays, where successive convex approximation (SCA) and semidefinite relaxation (SDR) are leveraged to handle non-convex constraints. Furthermore, the minimum number of dedicated sensing beams is analyzed via a worst-case rank bound, upon which the proposed beamforming design is further extended to the hybrid analog-digital (HAD) array architecture, where the unit-modulus constraint is addressed by manifold optimization. Simulation results demonstrate that only a small number of sensing beams are sufficient for both sensing and jamming AEs, and the proposed designs consistently outperform strong baselines while also revealing the communication–sensing trade-off.}

\end{abstract}


\IEEEpeerreviewmaketitle

\section{Introduction}
In recent years, low-altitude communication networks have attracted considerable attention, which provides flexible and dynamic support for the 5th-generation and beyond communication systems \cite{11006083}. However, the constituent devices in such networks, such as unmanned aerial vehicles (UAVs), also pose significant security risks when acting as eavesdroppers. Due to high mobility, UAVs can position themselves to avoid obstructions and establish strong line-of-sight (LoS) channels, which complicates secure communication. Physical-layer security (PLS) is expected to mitigate eavesdropping threats by exploiting differences between the legitimate and eavesdropping links \cite{10054167}. A key challenge is obtaining the eavesdropper’s channel state information (CSI), which is difficult because eavesdroppers are typically passive and non-cooperative \cite{10375133}.

Integrated sensing and communication (ISAC) offers a principled way to address this challenge. The ISAC technique effectively reduces hardware costs and improves spectrum efficiency by combining sensing and communication functionalities into a single system \cite{10086626}. Furthermore, the two functionalities can be co-designed to achieve mutual benefits rather than being viewed as separate objectives \cite{9737357}. Especially, echoes of ISAC signals enable the estimation of delay, Doppler, and angle of arrival (AoA), facilitating CSI reconstruction. For instance, the location and AoA have been estimated to aid beam alignment in \cite{7888145,8835615}. Furthermore, an extended Kalman filtering (EKF) framework was proposed in \cite{9171304} to predict the sensing parameters, thus improving the beam tracking accuracy in high-mobility communication scenarios. In \cite{10304580}, deep-learning-based methods were employed to extract the CSI of communication users in a reconfigurable intelligent surface (RIS)-assisted downlink ISAC system. \textcolor{black}{Inspired by these works, eavesdroppers' CSI can likewise be inferred via sensing to enhance secrecy design.} Note that to achieve high sensing accuracy, the ISAC signals should be steered towards the target to enhance the estimation accuracy for eavesdroppers' parameters, which, however, increases the risk of information leakage \cite{9593096}.

One possible solution to the above issue is to superimpose extra dedicated pseudo-random sensing signals onto the confidential communication signals \cite{6848758,9838753}. \textcolor{black}{Such sensing signals can also serve as artificial noise (AN), degrading eavesdropping channels while improving sensing accuracy.} For such cases, the authors in \cite{10153696} jointly designed transmit communication and sensing waveforms to minimize the beampattern matching mean squared error (MSE) under secure communication constraints. 
In \cite{10605793}, a joint design of transmit beamforming was proposed for a secure cell-free ISAC system to maximize the target detection probability against both information and sensing eavesdropping. When the target location is unknown, the authors in \cite{10227884} proposed a two-stage beamforming design scheme, in which an omnidirectional waveform is first emitted to sense the eavesdropper's direction, and then the communication beamforming matrix and AN matrix are optimized to maximize the secrecy rate and minimize the Cramér-Rao Bound (CRB) for the eavesdropper's direction estimation. As a step further, in \cite{10639496}, the sensing posterior CRB (PCRB) was derived with the eavesdropper's probability mass function, based on which the transmitted waveform was designed against multiple possible locations of the eavesdropper.

Most existing studies adopted fully digital arrays to transmit the same number of sensing beams as the number of antennas \cite{10050406,10364735}. In this case, the sensing covariance matrix is full-rank and can thus be efficiently optimized via semidefinite programming (SDP), followed by eigenvector decomposition to derive the sensing beamforming vectors. However, fully digital arrays may be impractical in massive multiple-input multiple-output (MIMO) systems due to their high hardware cost. Therefore, the hybrid analog-digital (HAD) array architecture was investigated to achieve beamforming with limited radio frequency (RF) chains, which are connected to massive antennas via low-cost phase shifters. 

Although it was shown in \cite{6966076} that the HAD beamforming can match the performance of fully digital beamforming when the number of RF chains is twice the number of data streams and pseudo-random sequences, in practice, to reduce hardware costs, the number of RF chains may be further limited. \textcolor{black}{This motivates studying the minimum number of sensing beams needed, as considered in recent works}. In \cite{10619398}, the authors derived that at most one sensing beam is needed for PCRB minimization in a single-target single-user ISAC system. Thereafter, the authors in \cite{chen2020composite} investigated the radar Signal-to-Interference-plus-Noise Ratio (SINR) maximization problem under the users' SINR constraints, which reveals that adding one dedicated sensing beam can significantly improve sensing performance. Furthermore, the authors in \cite{10942665} explored the minimum number of sensing beams required for estimating a given number of variables. In \cite{yao2025optimal}, the authors derived a more general bound on the number of sensing beams for multi-antenna multi-target multi-user ISAC systems. 

All the aforementioned studies demonstrated that using a limited number of sensing beams can still achieve excellent ISAC system performance. However, the dedicated sensing signals in secure ISAC systems serve dual purposes, i.e., jamming and sensing eavesdroppers. It remains unclear whether only a few sensing beams are sufficient for both roles. Moreover, how to design beamforming matrices for dedicated sensing signals with highly limited RF chains in an HAD array while maintaining satisfactory performance remains an open question. Motivated by the above, we study the secure downlink transmission with multiple users and aerial eavesdroppers (AEs) for both fully digital and HAD arrays. First, we design beamforming for the fully digital architecture and analyze a worst-case rank bound that upper-bounds the minimum number of sensing beams without degrading sensing. Second, the beamforming design is extended to the HAD array to reduce the hardware cost. Our main contributions are summarized as follows.

\begin{figure}[t]
	\centering
	\includegraphics[width=0.9\linewidth]{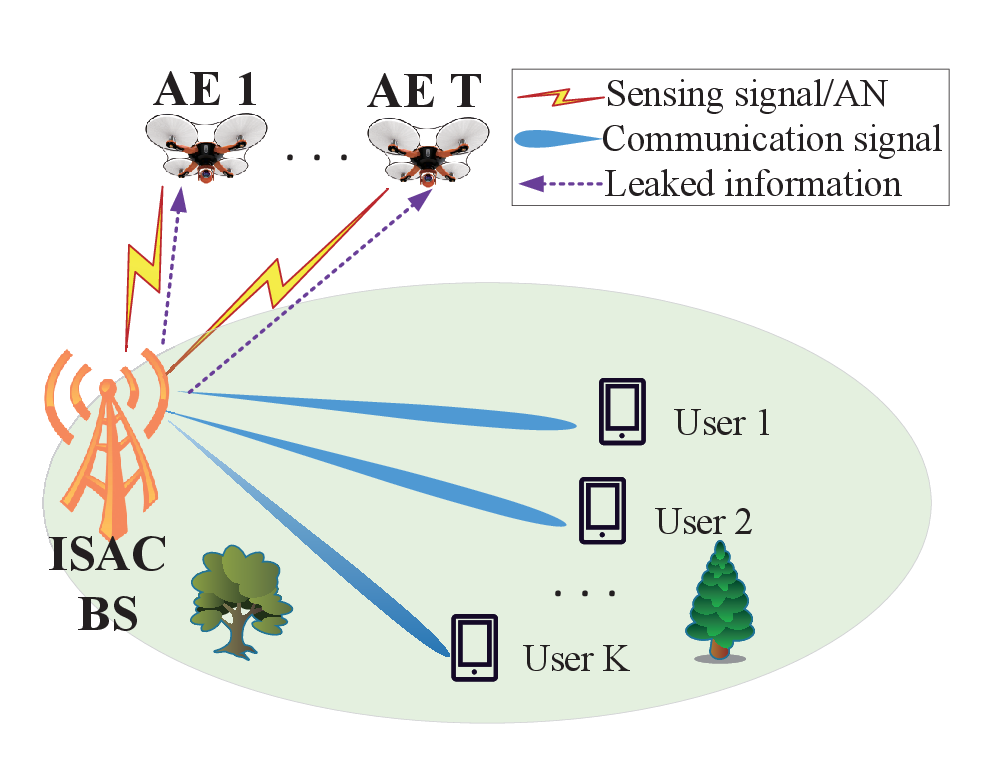}
	\caption{An illustration of the considered secure ISAC system.}
	\label{systemmodel}
\end{figure}

\begin{itemize}
\item We propose a novel sensing-assisted secure communication framework, where an ISAC base station (BS) transmits confidential communication signals and dedicated sensing signals to serve communication users against aerial eavesdroppers (AEs), as illustrated in Fig. \ref{systemmodel}. Therein, the sensing results for eavesdroppers are leveraged to facilitate beamforming design. A joint communication and sensing beamforming design problem is formulated to maximize the system sum rate, while constraining the AEs' SINR to guarantee secrecy performance and constraining the sensing Signal-to-Clutter-plus-Noise Ratio (SCNR) to guarantee sensing accuracy.

\item For fully digital arrays, we propose a fractional programming (FP)-based alternating beamforming optimization algorithm, where the successive convex approximation (SCA) and semidefinite relaxation (SDR) techniques are employed to address the non-convex constraints. Moreover, the monotonic convergence of the algorithm is proved, the initialization scheme of the proposed algorithm is provided, and the computational complexity is analyzed.

\item We establish a worst-case rank bound on the sensing covariance, which upper-bounds the minimum number of dedicated sensing beams required while ensuring the sensing and jamming performance.

\item We extend the design to the HAD array, where the digital beamforming matrix and analog beamforming matrix are optimized iteratively. In particular, we propose a penalty-based manifold optimization algorithm to optimize the analog beamforming matrix, where the unit-modulus constraints are addressed by manifold optimization, and the other constraints are addressed by treating them as penalty terms.


\end{itemize}

The remainder of this paper is organized as follows. Section II introduces the system model, the signal model, and the problem formulation. The FP-based alternating beamforming optimization is proposed for fully digital arrays in Section III. The HAD beamforming design is proposed in Section IV. Section V presents simulation results. Finally, Section VI concludes this paper.

\textit{Notations}: $a$, $\bf{a}$, and $\bf{A}$ denote complex scalar value, vector, and matrix, respectively; ${\left[  \cdot  \right]^*}$, ${\left[  \cdot  \right]^T}$, and ${\left[  \cdot  \right]^H}$ denote the conjugate, transpose, and conjugate-transpose operations, respectively; $\left| {\cdot} \right|$, $\left\|  \cdot  \right\|$, and $\left\|  \cdot  \right\|_F$ denote the absolute value, Euclidean norm, and Frobenius norm, respectively; $\mathbb{C}$ represents the set of complex numbers; $\Re \left\{ \cdot \right\}$ denotes the real part of its argument; ${\bf{I}}_n$ denotes the identity matrix with the dimension of $n\times n$; ${\bf{0}}_{m \times n}$ denotes the zero matrix with the dimension of $m\times n$;  $\mathcal{C} \mathcal{N}\left(\mu, \sigma^{2} \right)$ denotes the circularly symmetric complex Gaussian distribution with mean $\mu$ and variance $\sigma^{2}$; ${\bf{A}} \otimes {\bf{B}}$ represents the Kronecker product of ${\bf{A}}$ and ${\bf{B}}$; ${\bf{A}} \odot {\bf{B}}$ represents the Hadamard product of ${\bf{A}}$ and ${\bf{B}}$; ${\left[ \cdot \right]^ + }$ denotes $\max \left\{ {\cdot,0} \right\}$.

\section{System Model}	
As shown in Fig. \ref{systemmodel}, we consider a secure multi-user downlink ISAC system that comprises an ISAC BS equipped with $M_t$ transmit antennas and $M_r$ receive antennas, $K$ single-antenna communication users, and $T$ single-antenna AEs. The AEs hover near the BS to eavesdrop on the downlink information for the $K$ users. In particular, to enhance the sensing and secrecy performance, dedicated sensing signals are superimposed onto the communication signals, and the BS employs their echoes to track the $T$ AEs.\footnote{Before tracking the AEs, the BS first emits an omnidirectional waveform for detecting AEs. Thus, the number of AEs and the coarse direction information of each AE become available to the BS \cite{10227884}. Based on such prior sensing information, the beams are further designed to obtain a higher sensing SCNR.} The CSI of AEs is then reconstructed based on the estimated positions. The sets of users and AEs are denoted as $\mathcal{K} = \{1, 2, \cdots, K\}$ and $\mathcal{T} = \{1, 2, \cdots, T\}$, respectively.

\subsection{Communication Signal Model}
The BS transmits $K$ data streams and $N_s$ individual pseudo-random sequences to simultaneously enable secure communication and target sensing. The transmitted signal at time index $l$ is given by
\begin{align}
	\label{eq2.1}
    {\bf{x}}[l] = \sum\limits_{k\in \mathcal{K}} {{{\bf{w}}_k}{c_k}}[l]  + \sum\limits_{n\in \mathcal{N}_s} {{{\bf{v}}_n}{s_n}}[l]  = {\bf{Wc}}[l] + {\bf{Vs}}[l],
\end{align}
where ${\bf W}  = \left[{\bf w}_1, {\bf w}_2, \cdots, {\bf w}_K \right] \in \mathbb{C}^{M_t \times K}$ and ${\bf V}  = \left[{\bf v}_1, {\bf v}_2, \cdots, {\bf v}_{N_s} \right] \in \mathbb{C}^{M_t \times {N_s}}$ denote the communication beamforming matrix and sensing beamforming matrix, respectively, ${\bf{c}}\left[ l \right] = {\left[ {{c_1}\left[ l \right], \cdots ,{c_K}\left[ l \right]} \right]^T} \in \mathbb{C}^{K }$ denotes the transmitted data symbols for the $K$ users, and ${\bf{s}}[l] = {\left[ {{s_1}[l], \cdots ,{s_{{N_s}}}}[l] \right]^T} \in \mathbb{C}^{N_s}$ denotes the sensing signals. \textcolor{black}{The transmitted data symbols are assumed to be independent Gaussian random variables with zero mean and unit power, and the sensing signals are pseudo-random sequences.} The autocovariance matrix and cross-covariance matrices are expressed as ${{\bf{R}}_c} = \mathbb{E}\left\{ {{\bf{c}}[l]{{\bf{c}}^H[l]}} \right\} = {{\bf{I}}_K}$, ${{\bf{R}}_s} = \mathbb{E}\left\{ {{\bf{s}}[l]{{\bf{s}}^H[l]}} \right\} = {{\bf{I}}_{{N_s}}}$, and ${{\bf{R}}_{cs}} = \mathbb{E}\left\{ {{\bf{c}}[l]{{\bf{s}}^H[l]}} \right\} = {{\bf{0}}_{K \times {N_s}}}$, respectively. For notational convenience, we retain $l$ only if necessary in the following sections to simplify the expression. 

We denote the baseband channels between BS and user $k$ as ${\bf h}_k$, which follows the Saleh–Valenzuela (S-V) channel model comprising both LoS and non-line-of-sight (NLoS) paths, 
\begin{align}
	\label{eq2.2}
	{\bf h}_k = \sqrt{\rho d_{bk}^{-\alpha_0}} \left({\bf{a}}\left( {{\psi _k}} \right) + \sum\limits_{i = 1}^{{n_{NL}}} {{\alpha _{n,i}}{\bf{a}}\left( {{\psi _{n,i}}} \right)} \right),
\end{align}
where ${\rho}$ denotes the channel gain at the reference distance of 1 meter,  $d_{bk}$ denotes the distance between the BS and user $k$, $\alpha_0$ denotes the path loss exponent, $\psi_k$ denotes the angle of user $k$ with respect to the BS, $n_{NL}$ denotes the number of NLoS paths, $\alpha _{n, i}$ and $ \psi _{n, i} $ denote the small-scale fading and angle of the corresponding path, ${\bf{a}} ( \cdot )$ denotes the transmit steering vector of the BS,  
\begin{align}
	\label{eq2.3}
	{\bf{a}}\left( \psi  \right) = {\left[ {1,{e^{j\frac{{2\pi {d_c}}}{\lambda_c }\sin (\psi )}}, \cdots ,{e^{j\frac{{2\pi {d_c}}}{\lambda_c }(M_t - 1)\sin (\psi )}}} \right]^T},
\end{align} 
where $\lambda_c$ denotes the carrier wavelength, and $d_c$ denotes the antenna spacing, typically set as $d_c = 0.5 \lambda_c$. Note that the channels between the BS and users are assumed to be perfectly available based on the channel estimation.

Thus, the signal received by the $k$-th user is expressed as 
\begin{align}
	\label{eq2.4}
	\!\!\!\!{{{y}}_k} = {\bf{h}}_k^H{{\bf{w}}_k}{c_k} + \sum\limits_{j\in \mathcal{K}\backslash k} {{\bf{h}}_k^H{{\bf{w}}_j}{c_j}}  + \sum\limits_{n\in \mathcal{N}_s} {{\bf{h}}_k^H{{\bf{v}}_n}{s_n}}  + {{n}}_{k},
\end{align}
where ${{n}}_k \sim \mathcal{C} \mathcal{N}\left(0, \sigma_c^{2}  \right)$ denotes the additive white Gaussian noise at the $k$-th user. The SINR for user $k$ is given as 
\begin{align}
	\label{eq2.5}
	{\gamma _{c,k}} = \frac{{{{\left| {{\bf{h}}_k^H{{\bf{w}}_k}} \right|}^2}}}{{\sum\limits_{j\in \mathcal{K}\backslash k} {{{\left| {{\bf{h}}_k^H{{\bf{w}}_{j}}} \right|}^2}}  + \sum\limits_{n\in \mathcal{N}_s} {{{\left| {{\bf{h}}_k^H{{\bf{v}}_n}} \right|}^2}}  + {\sigma_c ^2}}}.
\end{align}

As for AEs, we consider the case where they maneuver in an open area. In this case, the channel from the BS to the $t$-th AE is dominated by the LoS component \cite{9857564}, which is expressed as
\begin{align}
	\label{eq2.6}
	{{\bf{h}}_{e,t}} = \sqrt {\rho d_{bt}^{ - 2}} {\bf{a}}\left( {{\theta _t}} \right),
\end{align}
where $d_{bt}$ denotes the distance between the BS and AE $t$, $\theta_t$ denotes the angle of AE $t$ with respect to the BS. Note that the channel from the BS to the $t$-th AE is difficult to obtain due to the non-cooperative nature. In this paper, the distances and angles of the AEs are assumed to be estimated by the BS via echo signals. Specifically, the distances are calculated using established methods such as matched-filtering \cite{richards2005fundamentals}, while the angles are estimated via algorithms like maximum likelihood estimation (MLE) or multiple signal classification (MUSIC) \cite{van2002optimum}. Subsequently, the channel ${{\bf{h}}_{e,t}}$ associated with the AE $t$ is reconstructed. In the following, the secure beamforming matrices are designed based on the reconstructed AEs' CSI. The received signal at AE $t$ is expressed as
\begin{align}
	\label{eq2.7}
	{{{y}}_{e,t}} =  \sum\limits_{k \in {\cal K}} {{\bf{h}}_{e,t}^H{{\bf{w}}_k}{c_k}}  + \sum\limits_{n \in {{\cal N}_s}} {{\bf{h}}_{e,t}^H{{\bf{v}}_n}{s_n}}  + {{n_t}},
\end{align}
where ${{n}}_t \sim \mathcal{C} \mathcal{N}\left(0, \sigma_e^{2}  \right)$ denotes the noise at the AE $t$.  The SINR for eavesdropping user $k$'s data stream at AE $t$ is written as\footnote{In this paper, we consider the worst case that the inter-device interference is perfectly eliminated using successive interference cancellation (SIC) at the AE \cite{7463025}, and secure beamforming is designed based on this assumption in the following section, ensuring communication security under other circumstances.}
\begin{align}
	\label{eq2.8}
	{\gamma _{e,t,k}} = \frac{{{{\left| {{\bf{h}}_{e,t}^H{{\bf{w}}_k}} \right|}^2}}}{{\sum\limits_{n\in \mathcal{N}_s} {{{\left| {{\bf{h}}_{e,t}^H{{\bf{v}}_n}} \right|}^2}}  + {\sigma_e ^2}}}.
\end{align}

Accordingly, the sum secrecy rate is expressed as
\begin{align}
	\label{eq2.8.1}
	\!\!\!\!{R_{\sec }} = {\sum\limits_{k \in \mathcal{K}} {\left[ {{{\log }_2}\left( {1 + {\gamma _{c,k}}} \right) - {{\log }_2}\left( {1 + \mathop {\max }\limits_{t \in \mathcal{T}} {\gamma _{e,t,k}}} \right)} \right]}^+}.
\end{align}

\subsection{Sensing Model}
The downlink communication signals and dedicated sensing signals can both be used to sense AEs. We define the combined beamforming matrix and combined symbol vector as ${{\bf{W}}_c} = \left[ {{{\bf{W}}},{{\bf{V}}}} \right]\in \mathbb{C}^{M_t \times \left(K+N_s\right) }$ and ${{\bf{s}}_c}[l] = {\left[ {{{\bf{c}}^T}[l],{{\bf{s}}^T}[l]} \right]^T} \in \mathbb{C}^{K+N_s} $, respectively. ${\bf{S}}_c = \left[ {{\bf{s}}_c[1], \cdots,{\bf{s}}_c[L]} \right] \in {\mathbb{C}^{(K+N_s)\times L}}$ denotes the transmitted signals in a Range-Doppler bin with the length being $L$. The echo signals received at BS over $L$ samples are expressed as 
\begin{align}
	\label{eq2.9}
	{{\bf{Y}}_s} = \sum\limits_{t = 1}^T {{{\bf{H}}_{s,t}}{{\bf{W}}_c}{{\bf{S}}_c}}  + {{\bf{H}}_I} {{\bf{W}}_c}{{\bf{S}}_c} + {\bf{N}},
\end{align}
where the first term denotes the desired echo signals, the second term is undesired echo signals from clutter, and the third term is the noise at the BS with ${\bf N}  \in \mathbb{C}^{M_r \times L}$ whose elements follow a Gaussian distribution with zero mean and variance $\sigma_s^2$.  Specifically, ${{\bf{H}}_{s,t}}$ denotes the BS-AE $t$-BS link, which is expressed as 
\begin{align}
	\label{eq2.10}
	{{\bf{H}}_{s,t}} = \sqrt {{\zeta _t}\rho d_{bt}^{ - 4}} {\bf{a}}\left( {{\theta _t}} \right){{\bf{a}}^H}\left( {{\theta _t}} \right),
\end{align}
where ${\zeta _t}$ denotes radar cross-section (RCS) of AE $k$. ${{\bf{H}}_I} = \sum\limits_{i = 1}^I {{\xi _i}{\bf{a}}({\theta _i}){{\bf{a}}^H}({\theta _i})}$ denotes the interference link, where  $\xi_i$ and $\theta_{i}$ denote the reflection coefficients and angle of the $i$-th clutter, respectively, $I$ denotes the number of clutters. The received signals are first matched filtered with the transmitted symbols to improve the sensing SCNR, i.e.,

\begin{equation}	
	\label{eq2.11}	
	\begin{aligned}	
	{{{\bf{\tilde Y}}}_s} &= {{\bf{Y}}_s}{\bf{S}}_c^H\\
	& = \sum\limits_{t = 1}^T{{\bf{H}}_{s,t}}{{\bf{W}}_c}{{\bf{S}}_c}{\bf{S}}_c^H + {{\bf{H}}_I}{{\bf{W}}_c}{{\bf{S}}_c}{\bf{S}}_c^H + {\bf{NS}}_c^H\\
	& = L\sum\limits_{t = 1}^T{{\bf{H}}_{s,t}}{{\bf{W}}_c} + L{{\bf{H}}_I}{{\bf{W}}_c} + {\bf{NS}}_c^H.
	\end{aligned} 
\end{equation}
Then, $T$ receive beamforming vectors, i.e., ${\bf u}_1,\cdots,{\bf u}_T$, are adopted to combine the signal in multiple antennas and separate the echoes from different AEs. The processed echoes corresponding to AE $t$ are expressed as
\begin{align}
	\label{eq2.12}
{\bf{y}}_{s,t}^H = L \sum\limits_{t = 1}^T{{\bf{u}}_t^H}{{\bf{H}}_{s,t}}{{\bf{W}}_c} + L{{\bf{u}}_t^H}{{\bf{H}}_I}{{\bf{W}}_c} + {{\bf{u}}_t^H}{\bf{NS}}_t^H,
\end{align}
where ${{\bf{u}}_t^H}{\bf{u}}_t = 1$. The set of receive beamforming vectors is expressed as ${\bf{U}} = \left[ {{{\bf{u}}_1}, \cdots ,{{\bf{u}}_T}} \right]$ in the following. The SCNR of the echoes from AE $t$ is expressed as
\begin{align}
	\label{eq2.13}
	{\gamma _{r,t}}& = \frac{{{{\left\| {L{\bf{u}}_t^H{{\bf{H}}_{s,t}}{{\bf{W}}_c}} \right\|}^2}}}{{{{\left\| {L{\bf{u}}_t^H{{\bf{H}}_I}{{\bf{W}}_c}} \right\|}^2} + {{\left\| {{\bf{u}}_t^H{\bf{NS}}_t^H} \right\|}^2}}}\\
	& = \frac{{L{\bf{u}}_t^H{{\bf{H}}_{s,t}}\left( {{\bf{W}}{{\bf{W}}^H} + {\bf{V}}{{\bf{V}}^H}} \right){\bf{H}}_{s,t}^H{{\bf{u}}_t}}}{{L{\bf{u}}_t^H{{\bf{H}}_I}\left( {{\bf{W}}{{\bf{W}}^H} + {\bf{V}}{{\bf{V}}^H}} \right){\bf{H}}_I^H{{\bf{u}}_t} + (K+N_s){\sigma_s ^2}}}.\notag
\end{align} 

The locations of AEs typically do not change significantly between two adjacent coherent time blocks. Hence, the sensing information from previous blocks is sufficient for facilitating the sensing SCNR calculation and for the following beamforming design \cite{9652071,10050406}.

\subsection{Problem Formulation}
One typical design criterion in secure transmission system is to maximize the sum secrecy rate defined in \eqref{eq2.8.1}. However, the presence of the term $\mathop {\max }\limits_{t \in \mathcal{T}} {\gamma _{e,t,k}}$ renders the objective function non-differentiable and introduces worst-case coupling among the AEs, which makes the problem highly non-convex and prevents the direct application of standard convex optimization techniques. Facing this challenge, we alternatively maximize the users' sum rate while constraining the maximum allowable AEs' SINR. This approach effectively enlarges the quality-of-service (QoS) gap between the users and AEs, thereby improving the secrecy capacity \cite{10054167,11105450}. Moreover, we constrain the SCNR of sensing to ensure the accuracy of reconstructed channels. The problem is formulated as\footnote{Maximizing the $k$-th user's achievable rate  $ \log_2\left(1 + \gamma_{c,k}\right)$ is equivalent to maximizing the natural logarithm $ \log\left(1 + \gamma_{c,k}\right)$. For convenience, we adopt the latter as the objective function to simplify derivation and computation.}
\begin{align}
	\label{eq2.14}
	\max _{{\bf{W}},{\bf{V}}, {\bf U}}\quad  &\sum\limits_{k \in {\cal K}} {{{\log }}\left( {1 + {\gamma _{c,k}}} \right)}  \\
	~{\rm {s.t.}}~\quad   &{\gamma _{e,t,k}} \le {\gamma _{e}} , \forall t, \forall k,  \tag{\ref{eq2.14}a}\\
	 &{\gamma _{r,t}} \ge {\gamma _{r}}, \forall t, \tag{\ref{eq2.14}b}\\
	& {\left\| {\bf{W}} \right\|_F^2} +{\left\| {\bf{V}} \right\|_F^2}\le P, \tag{\ref{eq2.14}c} 
\end{align}
where $\gamma_e$ denotes the maximum allowable AEs' SINR, ${\gamma _{r}}$ denotes the sensing SCNR requirement, and $P$ denotes the power budget.\footnote{Note that the value of ${\gamma _{r}}$ involves a trade-off and should be carefully chosen. A sufficiently high value ensures accurate CSI reconstruction based on reliable sensing; however, setting it too high may divert excessive resources to sensing, thus degrading communication performance. The communication-sensing trade-off is illustrated by the simulation results.} As can be seen, problem \eqref{eq2.14} is highly non-convex and difficult to solve due to the following reasons. On the one hand, the objective function and the constraints are non-convex due to the fractional terms, logarithmic functions, and the coupled variables. On the other hand, the number of constraints is large due to multiple users and multiple AEs, resulting in a non-convex and narrow feasible region. 

\section{Beamforming Design for Full-Digital Array}
In this section, a fully digital array architecture is considered to draw the important insights between the number of dedicated sensing beams and the secure communication performance. Moreover, the minimum required number of dedicated sensing beams is investigated. Specifically, we propose an alternating-optimization-based digital beamforming design algorithm. We first reformulate the problem into a more tractable form via the closed-form FP. Then, the problem is decomposed into three sub-problems, and the corresponding variables are optimized in an alternating manner using the SCA and SDR techniques.

\subsection{FP-based Transformation}
To deal with the fractional form and the logarithm functions, we convert the objective function \eqref{eq2.14} by Lagrangian dual transform \cite{8314727}. The data rate of the $k$-th device is rewritten as  
\begin{equation}	
	\label{eq3.1}	
	\begin{aligned}	
		& \log(1+\gamma_{c,k}) =   \max _{{\bf W}, {\bf V},\nu _k }~  \log  \left( {1  + {\nu _k}} \right)  - {\nu _k} \\
		&\quad \quad  + \left( {1 + {\nu _k}} \right) \frac{{{{\left| {{\bf h}_k^H{{\bf w}_k}} \right|}^2}}}{{\sum\limits_{j \in  {\cal K}} {{{\left| {{\bf h}_k^H{{\bf w}_j}} \right|}^2}}  +  \sum\limits_{n\in \mathcal{N}_s} {{{\left| {{\bf{h}}_k^H{{\bf{v}}_n}} \right|}^2}}+ \sigma_c^2}},
	\end{aligned} 
\end{equation}
where $\nu_k$ is the auxiliary variable. Maximizing the data rate for any given ${\nu _k}$ is equivalent to maximizing the fractional term $\frac{{{{\left| {{\bf h}_k^H{{\bf w}_k}} \right|}^2}}}{{\sum\limits_{j \in  {\cal K}} {{{\left| {{\bf h}_k^H{{\bf w}_j}} \right|}^2}} + \sum\limits_{n\in \mathcal{N}_s} {{{\left| {{\bf{h}}_k^H{{\bf{v}}_n}} \right|}^2}} + \sigma_c^2}}$. By applying the quadratic transform, problem \eqref{eq3.1} is reformulated as
\begin{align}
	\label{eq3.2}
	\max _{{{\bf{W}},{\bf V}, \beta_k}}~ & 2\sqrt {1 + {\nu _k}} {\mathop{\Re}\nolimits} \left( {\beta _k^*{\bf h}_k^H{{\bf w}_k}} \right) \\
	& - {\left| {{\beta _k}} \right|^2}\left( {\sum\limits_{j \in {\cal K}} {{{\left| {{\bf h}_k^H{{\bf w}_j}} \right|}^2}} +\sum\limits_{n\in \mathcal{N}_s} {{{\left| {{\bf{h}}_k^H{{\bf{v}}_n}} \right|}^2}} + \sigma_c^2} \right),\notag
\end{align}
where $\beta_k$ is the auxiliary variable. Based on the aforementioned transformation, the original objective function \eqref{eq2.14} is formulated as \eqref{eq3.3}, which is presented at the top of this page. In \eqref{eq3.3}, we define ${{\bf{R}}_v} = {\bf{V}}{{\bf{V}}^H}$, where ${{\bf{R}}_v} \ge 0$, ${\rm{rank}}\left( {{{\bf{R}}_v}} \right) = {N_s}$. The other variables are denoted as ${\boldsymbol \nu} = [\nu _1, \cdots, \nu _k]^T$, ${\boldsymbol \beta} = [\beta _1, \cdots, \beta _k]^T$, ${\cal N} = {\rm{diag}}\left( {2\sqrt {(1 + {\boldsymbol{\nu }})} } \right)$, ${\bf H} = \left[{\bf  h}_1, \cdots, {\bf h}_K \right]$, ${\bf \tilde  H} = {\bf H} {\rm diag}\left(\boldsymbol \beta \right) $.

\begin{figure*}[t]	
	\begin{equation}	
		\label{eq3.3}	
		\begin{aligned}	
			&r_{\rm sum} \left( \boldsymbol \nu,\boldsymbol \beta, {\bf W} ,{\bf R}_v \right) \\
			&= \sum\limits_{k \in  {\cal K}} \log \left( {1 + {\nu _k}} \right) - \sum\limits_{k \in  {\cal K}} {\nu _k} + \sum\limits_{k \in  {\cal K}} 2\sqrt {1 + {\nu _k}} {\mathop{\Re}\nolimits} \left( {\beta _k^*{\bf h}_k^H{{\bf w}_k}} \right) - \sum\limits_{k \in  {\cal K}} {\left| {{\beta _k}} \right|^2}\left( {\sum\limits_{j \in {\cal K}} {{{\left| {{\bf h}_k^H{{\bf w}_j}} \right|}^2}}  +\sum\limits_{n\in \mathcal{N}_s} {{{\left| {{\bf{h}}_k^H{{\bf{v}}_n}} \right|}^2}} + \sigma_c^2} \right) \\
			&= {\bf 1}_K^T \left( \log \left( {{\bf{1}}_K + {\boldsymbol \nu}} \right) -  {\boldsymbol \nu } \right)  + {\mathop{\Re}\nolimits} \left( {{\rm Tr}\left( {{{\mathcal N} {\bf \tilde H}^H {\bf W}}} \right)} \right) - {\left\| {{\bf{\tilde H}}^H {\bf W}} \right\|_F^2} -{\rm Tr}\left( {{{{\bf{\tilde H}}}^H}{{\bf{R}}_v}{\bf{\tilde H}}} \right) - {\left\| \boldsymbol \beta  \right\|^2}\sigma_c^2,
		\end{aligned} 
	\end{equation}
	\hrulefill
\end{figure*}

\subsection{Transformation for AEs' SINR Constraints}
To address the AEs' SINR constraints (\ref{eq2.14}a), we rewrite them by substituting ${{\bf{R}}_v} = {\bf{V}}{{\bf{V}}^H}$, yielding 
\begin{align}
	\label{eq3.4}
    \frac{1}{\gamma_e }{\bf{w}}_k^H{{\bf{h}}_{e,t}}{\bf{h}}_{e,t}^H{{\bf{w}}_k} - {\bf{h}}_{e,t}^H{{\bf{R}}_v}{{\bf{h}}_{e,t}} \le {\sigma_e ^2}, \forall t, \forall k,
\end{align} 
Notably, the reformulated constraints in \eqref{eq3.4} exhibit convexity.

\subsection{Transformation for Sensing SCNR Constraints}
We first introduce variables ${{\bf{A}}_{{\bf{H}},t}} = {\bf{H}}_{s,t}^H{{\bf{u}}_t}{\bf{u}}_t^H{{\bf{H}}_{s,t}}$, $\forall t \in \mathcal{T}$, and  ${{\bf{A}}_{{\bf{H}},I}} = {\bf{H}}_I^H{{\bf{u}}_t}{\bf{u}}_t^H{{\bf{H}}_I}$. Then the sensing SCNR constraint (\ref{eq2.14}b) is converted as 
\begin{align}
	\label{eq3.5}
&{\rm Tr}\left( {{{\bf{A}}_{{\bf{H}},t}}{\bf{W}}{{\bf{W}}^H}} \right) - {\rm Tr}\left( {{\gamma _r}{{\bf{A}}_{{\bf{H}},I}}{\bf{W}}{{\bf{W}}^H}} \right) \\
&\qquad \qquad+ {\rm Tr}\left( {\left( {{{\bf{A}}_{{\bf{H}},t}} - {\gamma _r}{{\bf{A}}_{{\bf{H}},I}}} \right){{\bf{R}}_v}} \right) \ge \frac{{{\gamma _r (K+N_s)}}}{L}{\sigma_s ^2}, \notag
\end{align} 
which is non-convex due to the quadratic term ${\rm Tr}\left( {{{\bf{A}}_{{\bf{H}},t}}{\bf{W}}{{\bf{W}}^H}} \right)$. By applying the first-order Taylor series expansion on $\bf{\tilde W}$, we obtain the following inequality
\begin{align}
	\label{eq3.6}
	&{\rm Tr}\left( {{{\bf{A}}_{{\bf{H}},t}}{\bf{W}}{{\bf{W}}^H}} \right) \\
	&\quad \ge 2\Re\left( {{\rm Tr}\left( {{{\bf{A}}_{{\bf{H}},t}}{\bf{\tilde W}}{{\bf{W}}^H}} \right)} \right) - {\rm Tr}\left( {{{\bf{A}}_{{\bf{H}},t}}{\bf{\tilde W}}{{{\bf{\tilde W}}}^H}} \right), \notag
\end{align} 
where $\bf{\tilde W}$ denotes the solution of $\bf W$ obtained in the previous iteration. Thus, the sensing SCNR constraint is reformulated as 
\begin{align}
	\label{eq3.7}
&2\Re\left( {{\rm Tr}\left( {{{\bf{A}}_{{\bf{H}},t}}{\bf{\tilde W}}{{\bf{W}}^H}} \right)} \right) - {\rm Tr}\left( {{\gamma _r}{{\bf{A}}_{{\bf{H}},I}}{\bf{W}}{{\bf{W}}^H}} \right) \notag \\
&\quad + {\rm Tr}\left( {\left( {{{\bf{A}}_{{\bf{H}},t}} - {\gamma _r}{{\bf{A}}_{{\bf{H}},I}}} \right){{\bf{R}}_v}} \right)   \\
&\ge \frac{{{\gamma _r (K+N_s)}}}{L}\sigma_s^2  + {\rm Tr}\left( {{{\bf{A}}_{{\bf{H}},t}}{\bf{\tilde W}}{{{\bf{\tilde W}}}^H}} \right), \forall t  \notag.
\end{align} 

\subsection{Alternating-Optimization for Fully Digital Array}
With the aforementioned transformation, problem \eqref{eq2.14} is rewritten as 
\begin{align}
	\label{eq3.8}
	\max _{\boldsymbol \nu,\boldsymbol \beta, {\bf{W}}, {\bf R}_v, {\bf{U}}}~  & r_{\rm sum} \left( \boldsymbol \nu,\boldsymbol \beta, {\bf W} ,{\bf R}_v \right)  \\
	~{\rm {s.t.}}\qquad     & \left\| {\bf{W}} \right\|_F^2 + {\rm{Tr}}\left( {{{\bf{R}}_v}} \right) \le P, \tag{\ref{eq3.8}a}\\
	& {\bf R}_v \succeq 0, {\rm rank}\left({\bf R}_v\right)  \le  N_s, \tag{\ref{eq3.8}b}\\
	& \eqref{eq3.4}, \eqref{eq3.7}. \notag
\end{align}
By decomposing problem \eqref{eq3.8}, the variables ${\boldsymbol \nu,\boldsymbol \beta, {\bf{W}}, {\bf R}_v}$ are decoupled and optimized alternately.
\subsubsection{Optimization of Auxiliary Variables $\boldsymbol \nu$ and $\boldsymbol \beta$} 
With other variables fixed, the optimal $\nu_k^\star$  is calculated by setting $\frac{{\partial {r_{sum}}}}{{\partial {\nu _k}}} = 0$, resulting in
\begin{align}
	\label{eq3.9}
	\nu_k^\star = \frac{{{{\left| {{\bf h}_k^H{{\bf w}_k}} \right|}^2}}}{{\sum\limits_{j \in  {\cal K}\backslash k} {{{\left| {{\bf h}_k^H{{\bf w}_j}} \right|}^2}} + {\bf{h}}_k^H{{\bf{R}}_v}{{\bf{h}}_k} + \sigma_c^2}}.
\end{align}
Similarly, the optimal $\beta_k^\star$ is obtained by setting $\frac{{\partial {r_{sum}}}}{{\partial {\beta _k}}} = 0$, resulting in
\begin{align}
	\label{eq3.10}
	\beta _k^ \star  = \frac{{\sqrt {1 + {\nu _k}} {\bf{h}}_k^H{{\bf{w}}_k}}}{{\sum\limits_{j \in {\cal K}} {{{\left| {{\bf{h}}_k^H{{\bf{w}}_j}} \right|}^2}}  + {\bf{h}}_k^H{{\bf{R}}_v}{{\bf{h}}_k} +  \sigma_c^2}}.
\end{align}

\subsubsection{Optimization of $\bf W$ and ${\bf R}_v$} 
We first drop the rank constraint on ${\bf R}_v$, and problem \eqref{eq3.8} is relaxed into an SDP problem and a Quadratic Programming (QP) problem with respect to ${\bf R}_v$ and $\bf W$, which can be optimized by using the interior-point method. 

\begin{theorem}
	(Worst-case Bound of Sensing beams): For the optimization problem \eqref{eq3.8} (or the original problem \eqref{eq2.14}), there exists an optimal solution ${{\bf{R}}_v^ \star }$ satisfying a worst-case rank bound, i.e.,
\begin{align}
	\label{eq3.11}
	{\rm{rank}}\left( {{\bf{R}}_v^ \star } \right) \le \sqrt{2T+1}.
\end{align}
\end{theorem}

\begin{IEEEproof} Please refer to Appendix 1.
\end{IEEEproof}
\textcolor{black}{This demonstrates that the optimal performance can be achieved by using no more than $\sqrt{2T+1}$ dedicated sensing beams. The required number of sensing beams scales with the square root of the number of AEs; thus, only a few sensing beams are sufficient to sense and jam the AEs in general. It is known from \cite{6966076} that the HAD beamforming can match the performance of fully digital beamforming when the number of RF chains is twice the number of digital beams. This reveals that adding dedicated sensing signals to the transmitted waveform can be applied to the HAD architecture without degrading the sensing and jamming performance. Considering $K$ data streams, the worst-case required number of RF chains is $2\left(\sqrt{2T+1}+K\right)$. In contrast, most existing works, such as \cite{10050406,10364735}, require sensing signals with a full-rank covariance matrix, which is feasible only for a fully digital architecture.}

\subsubsection{Optimization of Receive Beamforming Matrix $\bf U$}
It is readily seen that the best receive beamforming matrix achieves the highest sensing SCNR; hence, the optimization for ${\bf u}_t$ is equivalent to the following problem
\begin{align}
	\label{eq3.10.1}
	 \max _{ {\bf{u}_t}}~  &   \frac{{L{\bf{u}}_t^H{{\bf{H}}_{s,t}}\left( {{\bf{W}}{{\bf{W}}^H} + {\bf{V}}{{\bf{V}}^H}} \right){\bf{H}}_{s,t}^H{{\bf{u}}_t}}}{{L{\bf{u}}_t^H{{\bf{H}}_I}\left( {{\bf{W}}{{\bf{W}}^H} + {\bf{V}}{{\bf{V}}^H}} \right){\bf{H}}_I^H{{\bf{u}}_t} + (K+N_s){\sigma_s ^2}}} \\
	{\rm {s.t.}}~~    &  {{\bf{u}}_t^H}{\bf{u}}_t = 1, \tag{\ref{eq3.10.1}a}
\end{align}
which is a typical generalized Rayleigh quotient. By defining ${{\bf{G}}_t} = {{\bf{H}}_{s,t}}\left( {{\bf{W}}{{\bf{W}}^H} + {\bf{V}}{{\bf{V}}^H}} \right){\bf{H}}_{s,t}^H$, ${{\bf{R}}_t} = {{\bf{H}}_I}\left( {{\bf{W}}{{\bf{W}}^H} + {\bf{V}}{{\bf{V}}^H}} \right){\bf{H}}_I^H + {(K+N_s)}{\sigma_s ^2}{{\bf{I}}_{{M_t}}}/L$, and ${{\bf{D}}_t} = {\bf{R}}_t^{ - \frac{1}{2}}{{\bf{G}}_t}{\bf{R}}_t^{ - \frac{1}{2}}$, the optimal receive beamforming matrix ${\bf u}_t$ is the eigenvector corresponding to the largest eigenvalue of ${\bf D}_t$.

\subsection{Overall Algorithm and Complexity Analysis}
Based on the algorithms presented above, the overall alternating optimization algorithm for fully digital beamforming design is presented in Algorithm 1.

\textbf{Initialization:} 
For the proposed alternating optimization algorithm, the starting point has a significant impact on the convergence speed and overall performance. Here, we propose a heuristic initialization method. Intuitively, the communication beams are designed to align with the users for high communication capacity, i.e., ${{\bf{W}}_i} = \left[ {{{\bf{w}}_{i,1}}, \cdots,{{\bf{w}}_{i, K}}} \right]$, ${{\bf{w}}_{i,k}} =  {\bf{a}}\left( {{\psi _k}} \right)$.  Similarly, the dedicated sensing beams are designed to align with the AEs to sense their positions and interfere with AEs' eavesdropping,  i.e., ${{\bf{V}}_i} = \left[ {{{\bf{v}}_{i,1}}, \cdots,{{\bf{v}}_{i, T}}} \right]{{\bf{1}}_{T \times {N_s}}}$, ${{\bf{v}}_{i,k}} =  {\bf{a}}\left( {{\theta _t}} \right)$, where ${{\bf{1}}_{T \times {N_s}}}$ denotes the all-ones matrix with the dimension of $T \times {N_s}$, which is leveraged to transform the dimension of the sensing beamforming matrix for the case of $N_s \ne T$. To satisfy the power budget, the communication and sensing beamforming matrices are initialized as 
\begin{align}
	\label{eq3.12a}
	&{{\bf{W}}^{[0]}} = \sqrt P \frac{{\left( {1 - \mu } \right){{\bf{W}}_i}}}{{\left( {1 - \mu } \right)\left\| {{{\bf{W}}_i}} \right\|_F + \mu \left\| {{{\bf{V}}_i}} \right\|_F}},\\
	&{{\bf{V}}^{[0]}} = \sqrt P \frac{{\mu {{\bf{V}}_i}}}{{\left( {1 - \mu } \right)\left\| {{{\bf{W}}_i}} \right\|_F + \mu \left\| {{{\bf{V}}_i}} \right\|_F}}\label {eq3.12b},
\end{align}
where $\mu \in [0,1]$ denotes the power weighting factor, which is searched over 0 to 1 until the constraints are satisfied.

\textbf{Convergence Analysis:} 
\textcolor{black}{In Algorithm 1, the receiving beamforming matrix $\bf U$ is optimized to facilitate the satisfaction of the SCNR sensing constraints; in other words, the SCNR constraints are relaxed with optimized $\bf U$ when updating the other variables. The other variables $\boldsymbol \nu,\boldsymbol \beta, {\bf{W}}, {\bf R}_v$, are alternately updated to maximize the objective value of \eqref{eq3.8}. Therefore, the objective value is monotonically non-decreasing with each variable optimized. Moreover, since the objective value is upper bounded by a finite value, the algorithm is guaranteed to converge to a stationary point \cite{razaviyayn2013unified}.}

\begin{algorithm}[t]
	\caption{Proposed Beamforming Design Algorithm for Fully Digital Array}
	\begin{algorithmic}[1]
		\REQUIRE The communication beamforming matrix ${\bf W}^{[0]}$ and dedicated sensing beamforming matrix ${\bf V}^{[0]}$.
		\STATE Set iteration number $n=1$. 
		\REPEAT	
		\STATE Update $\boldsymbol \nu$ and $\boldsymbol \beta$ via \eqref{eq3.9} and \eqref{eq3.10}.
		\STATE Update $\bf W$ and ${\bf R}_v$ via interior-point method.
		\STATE Update $\bf U$ by solving problem \eqref{eq3.10.1}.
		\STATE Update $n=n+1$.
		\UNTIL {The objective value of \eqref{eq3.8} is converged or maximum iteration is reached.}
		\STATE Construct a low-rank covariance matrix of sensing beamforming vector via rank-reduction techniques in \cite{5233822}.
		\STATE Obtain the dedicated sensing beamforming matrix via eigenvalue decomposition.
		\ENSURE The optimized communication and sensing beamforming matrices $\bf W$ and $\bf V$.
	\end{algorithmic}
\end{algorithm}

\textbf{Computational Complexity Analysis:} The auxiliary variables $\boldsymbol \nu$ and $\boldsymbol \beta$ are updated using closed-form expressions, thus being computationally efficient. Consequently, the dominant computational complexity of the proposed algorithm stems from the updates of $\bf W$ and ${\bf R}_v$ with SDR technique, which is in the order of ${\mathcal{O}\left(N_\text{FP}M_t^{4.5} \log(1/\epsilon)\right)}$, where $N_\text{FP}$ denotes the number of FP iterations,  $\epsilon$ denotes the convergence threshold.

\begin{figure}[t]
	\centering
	\includegraphics[width=0.9\linewidth]{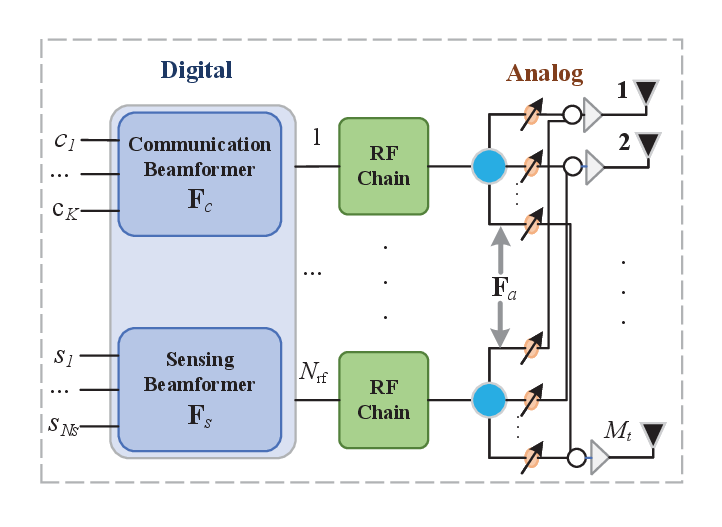}
	\caption{The HAD array architecture for transmitting both communication and dedicated sensing signals.}
	\label{HAD}
\end{figure}

\section{Beamforming Design for HAD Array with Limited RF Chains}
In the hybrid analog-digital (HAD) beamforming design, a widely-used approach is to decompose the fully digital beamforming matrix obtained in Section III \cite{7397861}. \textcolor{black}{Based on Theorem 1, $2\left(\sqrt{2T+1}+K\right)$ RF chains are required in the HAD architecture to achieve performance consistent with that of fully digital beamforming. However, in practice, employing fewer RF chains is desirable to reduce hardware cost, which unfortunately leads to severe performance degradation in the decomposition-based HAD beamforming scheme \cite{6966076}. To address this issue, we propose to directly optimize the analog beamforming matrix and digital beamforming matrix in an iterative manner with fewer RF chains.}

\subsection{HAD Array Architecture and Problem Formulation}
The HAD array architecture of the transmitter is illustrated in Fig. \ref{HAD}. The $K$ communication data streams and $N_s$ pseudo-random sequences are mapped to $N_{\rm{rf}}$ RF chains via a digital communication beamforming matrix ${{\bf{F}}_c} \in \mathbb{C}^{N_{\rm{rf}}\times K}$ and a digital sensing beamforming matrix ${{\bf{F}}_s} \in \mathbb{C}^{N_{\rm{rf}}\times N_s}$, respectively. Each RF chain is connected to $M_t$ transmit antennas through analog phase shifters. By utilizing an analog beamforming matrix ${{\bf{F}}_a} \in \mathbb{C}^{M_t \times N_{\rm{rf}}}$, the transmitted beams are steered. Therefore, the communication and sensing beamforming matrices are expressed as ${\bf{W}} = {{\bf{F}}_a}{{\bf{F}}_c}$ and ${\bf{V}} = {{\bf{F}}_a}{{\bf{F}}_s}$, respectively.  The transmitted signal of the BS with HAD array is written as
\begin{align}
	\label{eq4.1}
	\bf x &= \sum\limits_{k \in {\cal K}} {{{\bf{F}}_a}{{\bf{f}}_{c,k}}{c_k}}  + \sum\limits_{n \in {N_s}} {{{\bf{F}}_a}{{\bf{f}}_{s,n}}{s_n}} \\
	& = {{\bf{F}}_a}{{\bf{F}}_c}{\bf{c}} + {{\bf{F}}_a}{{\bf{F}}_s}{\bf{s}}, \notag
\end{align}
where ${{\bf{F}}_c} = \left[ {{{\bf{f}}_{c,1}}, \cdots {{\bf{f}}_{c,K}}} \right]$, ${{\bf{F}}_s} = \left[ {{{\bf{f}}_{s,1}}, \cdots {{\bf{f}}_{s,{N_s}}}} \right]$. Each element of analog beamforming matrix is unit-modulus, i.e., $\left| {{{\bf{F}}_{a, i,j}}} \right| = 1,\forall i,\forall j$, ${{{\bf{F}}_{a, i,j}}}$ denotes the $(i,j)$-th element of ${\bf F}_a$. Similarly, the receive beamforming matrix is expressed as ${\bf{U}} = {{\bf{U}}_a}{{\bf{U}}_d}$, where ${\bf{U}}_a$  denotes the analog receive beamforming matrix and ${\bf{U}}_d = \left[ {{{\bf{u}}_{d,1}}, \cdots {{\bf{u}}_{d, T}}} \right]$ denotes the digital receive beamforming matrix, respectively.

By substituting ${\bf{W}}$ and ${\bf{V}}$ with ${{\bf{F}}_a}{{\bf{F}}_c}$ and ${{\bf{F}}_a}{{\bf{F}}_s}$, we reformulate problem \eqref{eq3.8} as \eqref{eq4.2} at the top of next page. The constraint (\ref{eq4.2}a) is the AEs' SINR constraints. The  constraint (\ref{eq4.2}b) is the sensing SCNR constraint, where  ${{\bf{A}}_{{\bf{H}},t}} = {\bf{H}}_{s,t}^H{{\bf{U}}_a}{{\bf{u}}_{d,t}}{\bf{u}}_{d,t}^H{\bf{U}}_a^H{{\bf{H}}_{s,t}}$, and  ${{\bf{A}}_{{\bf{H}},I}} = {\bf{H}}_I^H{{\bf{U}}_a}{{\bf{u}}_{d,t}}{\bf{u}}_{d,t}^H{\bf{U}}_a^H{{\bf{H}}_I}$. The constraint (\ref{eq4.2}c) is the power budget constraint.

\begin{figure*}[t]	
	
	\begin{align}	
		\label{eq4.2}
		\mathop {\max }\limits_{{\bf{\nu }},{\bf{\beta }},{{\bf{F}}_a},{{\bf{F}}_c}\hfill\atop
			{{\bf{F}}_s},{{\bf{U}}_a},{{\bf{U}}_d}\hfill} \;\; & {{r}}_{HAD}  = {\bf{1}}_K^T\left( {{{\log }}\left( {{\bf{1}}_K + {\boldsymbol{\nu }}} \right) - {\boldsymbol{\nu }}} \right) + \Re \left( {{\rm Tr}\left( {{\cal N}{{{\bf{\tilde H}}}^H}{{\bf{F}}_a}{{\bf{F}}_c}} \right)} \right) - {\rm Tr}\left( {{{{\bf{\tilde H}}}^H}{{\bf{F}}_a}\left( {{{\bf{F}}_c}{\bf{F}}_c^H + {{\bf{F}}_s}{\bf{F}}_s^H} \right){\bf{F}}_a^H{\bf{\tilde H}}} \right) - {\left\| {\boldsymbol{\beta }} \right\|^2}\sigma_c^2, \\
		\mathrm{s.t.}\quad  ~  &\frac{1}{\gamma_e }{\bf{f}}_{c,k}^H{\bf{F}}_a^H{{\bf{h}}_{e,t}}{\bf{h}}_{e,t}^H{{\bf{F}}_a}{{\bf{f}}_{c,k}} - {\bf{h}}_{e,t}^H{{\bf{F}}_a}{{\bf{F}}_s}{\bf{F}}_s^H{\bf{F}}_a^H{{\bf{h}}_{e,t}} \le {\sigma_e ^2},\forall k,\forall t, \tag{\ref{eq4.2}a} \label{eq4.2a}\\
		& {\rm Tr}\left( {\left( {{{\bf{A}}_{{\bf{H}},t}} - {\gamma _r}{{\bf{A}}_{{\bf{H}},I}}} \right){{\bf{F}}_a}\left( {{{\bf{F}}_c}{\bf{F}}_c^H + {{\bf{F}}_s}{\bf{F}}_s^H} \right){\bf{F}}_a^H} \right) \ge \frac{{{\gamma _r (K+N_s)}}}{L}{\sigma_s ^2},\forall t,\tag{\ref{eq4.2}b} \label{eq4.2b}\\
		& {\rm Tr}\left( {{\bf{F}}_a^H{{\bf{F}}_a}\left( {{{\bf{F}}_c}{\bf{F}}_c^H + {{\bf{F}}_s}{\bf{F}}_s^H} \right)} \right) \le P, \tag{\ref{eq4.2}c} \label{eq4.2c}\\
		& \left| {{{\bf{F}}_{a,i,j}}} \right| = 1,\forall i,\forall j, \tag{\ref{eq4.2}d} \label{eq4.2d}
	\end{align}

	\hrulefill
\end{figure*}

\subsection{Alternating-Optimization for HAD Array}

\subsubsection{Optimization of Auxiliary Variables $\boldsymbol \nu$ and $\boldsymbol \beta$} 
Similar to the fully digital array setup, $\boldsymbol \nu$ and $\boldsymbol \beta$ are obtained from the stationarity condition, i.e., by setting  $\frac{{\partial {r_{HAD}}}}{{\partial {\nu _k}}} = 0$ and $\frac{{\partial {r_{HAD}}}}{{\partial {\beta _k}}} = 0$, resulting in
\begin{align}
	\label{eq4.3}
	\nu _k^ \star  = \frac{{{{\left| {{\bf{h}}_k^H{{\bf{F}}_a}{{\bf{f}}_{c,k}}} \right|}^2}}}{{{\sum \limits_{j \in K\backslash k}}{{\left| {{\bf{h}}_k^H{{\bf{F}}_a}{{\bf{f}}_{c,j}}} \right|}^2} + {{\left\| {{\bf{h}}_k^H{{\bf{F}}_a}{{\bf{F}}_s}} \right\|}^2} + \sigma_c^2}},
\end{align}
\begin{align}
	\label{eq4.4}
	\beta _k^ \star  = \frac{{\sqrt {1 + {\nu _k}} {\bf{h}}_k^H{{\bf{F}}_a}{{\bf{f}}_{c,k}}}}{{\sum\limits_{j \in K} {{{\left| {{\bf{h}}_k^H{{\bf{F}}_a}{{\bf{f}}_{c,j}}} \right|}^2}}  + {{\left\| {{\bf{h}}_k^H{{\bf{F}}_a}{{\bf{F}}_s}} \right\|}^2} + \sigma_c^2}}.
\end{align}

\subsubsection{Optimization of Digital Beamforming Matrix ${\bf F}_c$ and ${\bf F}_s$} 
With the other variables fixed, problem \eqref{eq4.2} is rewritten as
\begin{align}	
	\label{eq4.5}
	\mathop {\max }\limits_{{{\bf{F}}_c},{{\bf{F}}_s}} ~~& {{r}}_{HAD}   \\
	\mathrm{s.t.}\quad & \eqref{eq4.2a},\eqref{eq4.2b},\eqref{eq4.2c}, \notag
\end{align}
which is still non-convex due to the quadratic constraints (\ref{eq4.2}a) and (\ref{eq4.2}b). By using the SCA technique, we transform the constraints as
\begin{align}
	\label{eq4.6}
	&\frac{1}{{{\gamma _e}}}{\bf{f}}_{c,k}^H{\bf{F}}_a^H{{\bf{h}}_{e,t}}{\bf{h}}_{e,t}^H{{\bf{F}}_a}{{\bf{f}}_{c,k}} - 2\Re \left( {{\rm Tr}\left( {{\bf{F}}_a^H{{\bf{h}}_{e,t}}{\bf{h}}_{e,t}^H{{\bf{F}}_a}{{{\bf{\tilde F}}}_s}{\bf{F}}_s^H} \right)} \right)  \notag \\
	&\qquad + {\rm Tr}\left( {{\bf{F}}_a^H{{\bf{h}}_{e,t}}{\bf{h}}_{e,t}^H{{\bf{F}}_a}{{{\bf{\tilde F}}}_s}{\bf{\tilde F}}_s^H} \right) \le {\sigma_e ^2},\forall k,\forall t,
\end{align}
\begin{align}
	\label{eq4.7}
	&2\Re \left( {{\rm Tr}\left( {{{\bf{A}}_{{{\bf{F}}_a}}}\left( {{{{\bf{\tilde F}}}_c}{\bf{F}}_c^H + {{{\bf{\tilde F}}}_s}{\bf{F}}_s^H} \right)} \right)} \right) \\
	& \quad - {\rm Tr}\left( {{{\bf{A}}_{{{\bf{F}}_a}}}\left( {{{{\bf{\tilde F}}}_c}{\bf{\tilde F}}_c^H + {{{\bf{\tilde F}}}_s}{\bf{\tilde F}}_s^H} \right)} \right) \ge \frac{{{\gamma _r}{(K+N_s)}}}{L}{\sigma_s ^2}, \forall t, \notag
\end{align}
where ${{\bf{A}}_{{{\bf{F}}_a}}} = {\bf{F}}_a^H\left( {{{\bf{A}}_{{\bf{H}},t}} - {\gamma _r}{{\bf{A}}_{{\bf{H}},I}}} \right){{\bf{F}}_a}$ is the auxiliary matrix, $\bf{\tilde F}_c$ and $\bf{\tilde F}_s$ denotes the solution obtained in the previous iteration. By replacing the constraint \eqref{eq4.2a} and \eqref{eq4.2b} with \eqref{eq4.6} and \eqref{eq4.7}, problem \eqref{eq4.5} is a convex problem and can be efficiently solved via the interior point method. 

\subsubsection{Optimization of Analog Beamforming Matrix ${\bf F}_a$} 
We then optimize the analog beamforming matrix ${\bf F}_a$ with the other variables fixed. By applying the following matrix transformation 
\begin{align}
	\label{eq4.8.0}
	&{\rm Tr}\left( {{{\bf{P}}^H}{\bf{ZPQ}}} \right) = {\rm{vec}}{\left( {\bf{P}} \right)^H}\left( {{{\bf{Q}}^T} \otimes {\bf{Z}}} \right){\rm{vec}}\left( {\bf{P}} \right),\\
	&{\rm Tr}\left( {{{\bf{P}}^H}{\bf{Q}}} \right) = {\rm{vec}}{\left( {\bf{P}} \right)^H}{\rm{vec}}\left( {\bf{Q}} \right),
\end{align}
we recast problem \eqref{eq4.2} as the following more concise form
\begin{align}
	\label{eq4.8}
	\mathop {\max }\limits_{{{\bf{f}}_a}} ~&\Re \left( {{\bf{a}}_0^H{{\bf{f}}_a}} \right) - {\bf{f}}_a^H{{\bf{A}}_0}{{\bf{f}}_a}\\
	{\rm{s.t.}} ~~& {\bf{f}}_a^H{{\bf{B}}_m}{{\bf{f}}_a} - {b_m} \le 0, m = 1,\cdots,M_c,\tag{\ref{eq4.8}a} \label{eq4.8a} \\
	&\left| {{{\bf{f}}_{a,i}}} \right| = 1,\forall i, \tag{\ref{eq4.8}b} \label{eq4.8b}
\end{align}
where ${{\bf{f}}_a} = {\rm vec}({{\bf{F}}_a})$, ${{\bf{A}}_0} = {\left( {{{\bf{F}}_c}{\bf{F}}_c^H + {{\bf{F}}_s}{\bf{F}}_s^H} \right)^T} \otimes \left( {{\bf{\tilde H}}{{{\bf{\tilde H}}}^H}} \right)$, and ${{\bf{a}}_0} = {\rm{vec}}\left( {{\bf{\tilde H}}{{\cal N}^H}{\bf{F}}_c^H} \right)$. $M_c=(K + 1)T + 1$ denotes the number of constraints, ${\bf B}_m$ is defined in \eqref{eq4.9} at the top of next page, and $b_m$ is defined in \eqref{eq4.10}.
\begin{figure*}[t]
\begin{align}
	\label{eq4.9}
		{{\bf{B}}_m}=\begin{cases}
		{{{\left( {\frac{1}{\gamma_e }{{\bf{f}}_{c,k}}{\bf{f}}_{c,k}^H - {{\bf{F}}_s}{\bf{F}}_s^H} \right)}^T} \otimes {{\left( {{{\bf{h}}_{e,t}}{\bf{h}}_{e,t}^H} \right)}},\quad }  & {m = \left( {t - 1} \right)K + k},\\
		{ - {{\left( {{{\bf{F}}_c}{\bf{F}}_c^H + {{\bf{F}}_s}{\bf{F}}_s^H} \right)}^T} \otimes \left( {{{\bf{A}}_{{\bf{H}},t}} - {\gamma _r}{{\bf{A}}_{{\bf{H}},I}}} \right),}  &{m = KT + t},\\
		{{{\left( {{{\bf{F}}_c}{\bf{F}}_c^H + {{\bf{F}}_s}{\bf{F}}_s^H} \right)}^T} \otimes {{\bf{I}}_{{M_t} }},\quad } &{m = \left( {K + 1} \right)T + 1}.
	\end{cases} 
\end{align}
	\hrulefill
\end{figure*}
\begin{align}
	\label{eq4.10}
	\!\!\!\!\!\!\!\!	{b_m}=\begin{cases}
		{{\sigma_e ^2},\quad }  & {m = \left( {t - 1} \right)K + k},\\
		{ - \frac{{{\gamma _r}{(K+N_s)}}}{L}\sigma_s^2,}  & {m = KT + t},\\
	    {P,\quad } & {m = \left( {K + 1} \right)T + 1.}
	\end{cases} \!\!\!\!\!\!\!
\end{align}

As can be observed, problem \eqref{eq4.8} is a non-convex problem due to the quadratic constraint \eqref{eq4.8a} and unit-modulus constraint \eqref{eq4.8b}. A common technique for addressing the constraints is SDR \cite{5447068}. However, the optimization variable ${{\bf{f}}_a}$ is a high-dimensional vector of size $M_t N_{\rm{rf}}\times 1$. The computational complexity of using the SDR technique is relatively high, at ${\mathcal{O}\left(N_{\text{FP}}M_t^{4.5} N_{\rm{rf}}^{4.5} \log(1/\epsilon)\right)}$, where $N_{\text{FP}}$ denotes the number of FP iterations, and $\epsilon$ denotes the convergence threshold. Moreover, the SDR technique does not ensure a rank-one solution, thereby violating the original problem constraints. 

To address this issue, we propose a penalty-based manifold optimization algorithm to optimize the analog beamforming matrix. We first convert the constraints into penalty terms, and then integrate them into the objective function. As a result, problem \eqref{eq4.8} is reformulated as
\begin{align}
	\label{eq4.11}
	\mathop {\max }\limits_{{{\bf{f}}_a}} ~& G\left( {{{\bf{f}}_a}} \right) = \Re \left( {{{\bf{a}}^H}{{\bf{f}}_a}} \right) - {\bf{f}}_a^H{\bf{A}}{{\bf{f}}_a} - \sum\limits_{m = 1}^{M_c} {{\lambda _m}{\Phi _m}\left( {{{\bf{f}}_a}} \right)}\\
	{\rm{s.t.}} ~~&\left| {{{\bf{f}}_{a,i}}} \right| = 1,\forall i, \tag{\ref{eq4.11}a} \label{eq4.11a}
\end{align}
where ${\Phi _m}\left( {{{\bf{f}}_a}} \right) = {\left( {{{\left[ {{\bf{f}}_a^H{{\bf{B}}_m}{{\bf{f}}_a} - {b_m}} \right]}^ + }} \right)^2}$ denotes the penalty of the $m$-th constraint, $\lambda_m$ denotes the weight factor of the $m$-th penalty term, which increases with iterations to enforce the satisfaction of the constraints. Note that problem \eqref{eq4.11} is defined on a complex unit circle manifold, whose feasible region is ${\cal M} = \left\{ {{{\bf{f}}_a}|\left| {{{\bf{f}}_{a, i}}} \right| = 1} \right\}$, which can be addressed using manifold optimization techniques \cite{Absil2009Optimization}. Specifically, we first express the \textit{tangent space} at a point ${\bf f}_a$ of the manifold $\mathcal{M}$ as follows, which is the set of all the tangent vectors at $\bf t$,
\begin{align}
	\label{eq4.12}
	T_{{\bf f}_a}\mathcal{M} = \left\{ \mathbf{t} \in \mathbb{C}^{M_t N_{\rm{rf}} \times 1} \ \middle| \ \Re\left( {\bf f}_a \odot {\bf{t}}^* \right) = \mathbf{0}_{M_t N_{\rm{rf}} \times 1} \right\}.
\end{align}
The Riemannian gradient is defined as the direction of steepest ascent of the objective function within the tangent space \textit{tangent space} $T_{{\bf f}_a}\mathcal{M}$, which is obtained by projecting the Euclidean gradient onto the tangent space. The Euclidean gradient of $G\left( {{{\bf{f}}_a}} \right)$ in \eqref{eq4.11} is calculated as 
\begin{align}
	\label{eq4.13}
	\nabla G\left( {{{\bf{f}}_a}} \right) = {{\bf{a}}_0} - 2{{\bf{A}}_0}{{\bf{f}}_a} - \sum\limits_{m = 1}^{M_c} {{\lambda _m}\nabla {\Phi _m}\left( {{{\bf{f}}_a}} \right)},
\end{align}
where $\nabla {\Phi _m}\left( {{{\bf{f}}_a}} \right)$ is given as 
\begin{align}
	\label{eq4.14}
	\!\!\!\!\!\!\!\!	\nabla {\Phi _m}\left( {{{\bf{f}}_a}} \right)=\begin{cases}
		{4\left( {{\bf{f}}_a^H{{\bf{B}}_m}{{\bf{f}}_a} - {b_m}} \right){{\bf{B}}_m}{{\bf{f}}_a}}, \!\!\! & {{\bf{f}}_a^H{{\bf{B}}_m}{{\bf{f}}_a} > {b_m}},\\
		0,&\rm{otherwise}.
	\end{cases} \!\!\!\!\!\!\!
\end{align}

Based on the Euclidean gradient shown in \eqref{eq4.13}, the Riemannian gradient is derived as
\begin{align}
	\label{eq4.15}
	{\rm{grad}}G\left( {{{\bf{f}}_a}} \right) = \nabla G\left( {{{\bf{f}}_a}} \right) - \Re \left( {\nabla G\left( {{{\bf{f}}_a}} \right) \odot {\bf{f}}_a^*} \right) \odot {{\bf{f}}_a}.
\end{align}
Thereafter, the variable ${\bf f}_a$ is updated with the direction of the Riemannian gradient
\begin{align}
	\label{eq4.15.1}
	\mathbf{f}_a^{[n+1]} \leftarrow \mathcal{R}_{\mathbf{f}_a^{[n]}}\left( \mu \, \mathrm{grad}\, G(\mathbf{f}_a^{[n]}) \right),
\end{align}
where $n$ is the iteration index, $\mu$ is the step size, $\mathcal{R}_{\mathbf{f}_a^{[n]}}\left(\cdot \right)$ is the retraction mapping to ensure the updated point remains on the complex circle manifold $\mathcal{M}$, which is given by
\begin{align}
\!\!\!\!{{\cal R}_{{\bf{f}}_a^{[n]}}}\left( {\mu {\mkern 1mu} {\rm{grad}}{\mkern 1mu} G({\bf{f}}_a^{[n]})} \right) = \left[ {\frac{{{{\left( {{\bf{f}}_a^{[n]} + \mu {\mkern 1mu} {\rm{grad}}{\mkern 1mu} G({\bf{f}}_a^{[n]})} \right)}_i}}}{{\left| {{{\left( {{\bf{f}}_a^{[n]} + \mu {\mkern 1mu} {\rm{grad}}{\mkern 1mu} G({\bf{f}}_a^{[n]})} \right)}_i}} \right|}}} \right].
\end{align}
Consequently, the optimized analog beamforming ${\bf F}_a$ is obtained by reshaping ${\bf f}_a$.

\subsubsection{Optimization of Receive Beamforming Matrices ${\bf U}_a$ and ${\bf U}_d$} 
To optimize the analog receive beamforming matrix and digital receiver beamforming matrix, we first optimize the fully digital receiving beamforming matrix by solving problem \eqref{eq3.10.1}. Then, the hybrid receive beamforming matrices are obtained by decomposing the fully digital beamforming matrix \cite{7397861}.

\subsection{Overall Algorithm and Complexity Analysis}
Based on the aforementioned algorithm, we present the overall alternating optimization algorithm in Algorithm 2. The initialization of the variables is presented as follows.

\textbf{Initialization:} 
The initial hybrid beamforming matrix is obtained by approximating the initial fully digital beamforming matrix in \eqref{eq3.12a} and \eqref{eq3.12b}, i.e.,
\begin{align}
	\label{eq4.16}
	\!\!\!\mathop {\min }\limits_{{\bf{F}}_a^{[0]},{\bf{F}}_c^{[0]},{\bf{F}}_s^{[0]}}~ &\left\| {{{\bf{W}}^{[0]}} - {\bf{F}}_a^{[0]}{\bf{F}}_c^{[0]}} \right\|_F^2 + \left\| {{{\bf{V}}^{[0]}} - {\bf{F}}_a^{[0]}{\bf{F}}_s^{[0]}} \right\|_F^2\\
	\rm {s.t.}~~~ \quad&\left| {{\bf{F}}_{a,i,j}^{[0]}} \right| = 1,\forall i, \forall j, \tag{\ref{eq4.16}a} \label{eq4.16a}\\
	& \left\| {{\bf{F}}_a^{[0]}{\bf{F}}_c^{[0]}} \right\|_F^2 + \left\| {{\bf{F}}_a^{[0]}{\bf{F}}_s^{[0]}} \right\|_F^2 \le P, \tag{\ref{eq4.16}b} \label{eq4.16b}
\end{align}
where ${{\bf{F}}_a^{[0]},{\bf{F}}_c^{[0]},{\bf{F}}_s^{[0]}}$ denote the initial analog beamforming matrix, digital communication beamforming matrix, and digital sensing beamforming matrix, respectively. Problem \eqref{eq4.16} is effectively solved in \cite{7397861}. The penalty factor $\lambda_m$ is initialized to balance the objective value and penalty terms,
\begin{align}
	\label{eq4.17}
	\lambda _m^{[0]} = \frac{{\left| {\Re \left( {{{\bf{a}}^H}{\bf{f}}_a^{[0]}} \right) - {\bf{f}}_a^{[0]H}{\bf{Af}}_a^{[0]}} \right|}}{{\left| {{\Phi _m}\left( {{\bf{f}}_a^{[0]}} \right)} \right|}},
\end{align}
where ${\bf{f}}_a^{[0]} = {\rm vec}\left({\bf F}_a^{[0]}\right)$.

\textbf{Convergence Analysis:}
\textcolor{black}{The proposed algorithm alternates between optimizing the digital and analog beamforming matrices. For the digital beamforming matrix optimization, a sub-optimal solution is obtained via the SCA technique, which guarantees that the objective function of problem \eqref{eq4.2} is monotonically increased throughout the iterative process \cite{9868348}. For the analog beamforming matrix optimization, the search direction in the manifold is ensured to be an ascent direction, ensuring that the objective value is non-decreasing. Therefore, the proposed HAD beamforming design algorithm is guaranteed to converge.}

\begin{algorithm}[t]
	\caption{Proposed Alternating Beamforming Design Algorithm for HAD Array}
	\begin{algorithmic}[1]
		\REQUIRE The analog beamforming matrix ${\bf F}_a^{[0]}$, digital communication beamforming matrix ${\bf F}_c^{[0]}$, digital sensing beamforming matrix ${\bf F}_s^{[0]}$.
		\STATE Set iteration number $n=1$. 
		\REPEAT	
		\STATE Update $\boldsymbol \nu$ and $\boldsymbol \beta$ via \eqref{eq3.9} and \eqref{eq3.10}.
		\STATE Update digital beamforming matrix ${\bf F}_c$ and ${\bf F}_s$ by solving problem \eqref{eq4.5}.
		\REPEAT
		\STATE Compute Euclidean gradient $\nabla G$ via \eqref{eq4.13}.
		\STATE Compute Riemannian gradient ${\rm{grad}}G$ via \eqref{eq4.15}.
		\STATE Update ${\bf f}_a^{[n+1]}$ via \eqref{eq4.15.1}.
		\STATE Reshape ${\bf f}_a^{[n+1]}$ as the analog beamforming matrix ${\bf F}_a^{[n+1]}$.
		\STATE Update $n=n+1$.
		\UNTIL {The objective value of \eqref{eq4.11} is converged or maximum iteration is reached.}
		\STATE Update $\lambda_m=2\lambda_m,m = 1,\cdots,M_c$.
		\STATE Update receive beamforming matrices ${\bf U}_a$ and ${\bf U}_d$.
		\UNTIL {The sum rate in \eqref{eq3.8} is converged.}
		\ENSURE The optimized HAD beamforming matrix ${\bf F}_a$, ${\bf F}_c$, and ${\bf F}_s$.
	\end{algorithmic}
\end{algorithm}

\textbf{Computational Complexity Analysis:} The computational burden primarily arises from updating the digital beamforming matrix and the analog beamforming matrix. The digital beamforming matrix is solved by using the interior point method, which has a complexity of order $\mathcal{O}\left(N_{\rm{rf}}^{3.5}{(K + N_s)^{3.5}}\right)$. Meanwhile, the dominant complexity in optimizing the analog beamforming matrix comes from computing the gradient in \eqref{eq4.13}, which is $\mathcal{O}\left(M_c M_t^2 N_{\rm{rf}}^2\right)$. Therefore, the overall complexity is ${\mathcal{O}\left({N_{{\rm{FP}}}}\left( {N_{\rm{rf}}^{3.5}{{\left( {K + {N_s}} \right)}^{3.5}} + {N_{{\rm{iter}}}}{M_c}M_t^2N_{\rm{rf}}^2} \right)\right)}$, where $N_{{\rm{FP}}}$ denotes the number of FP iterations and $N_{{\rm{iter}}}$ denotes the number of iterations for manifold optimization.

\section{Simulation Results}
In this section, we validate the effectiveness of the proposed beamforming design schemes for secure downlink transmission.  Unless otherwise specified, we assume the ISAC BS serves $K=4$ communication users located at a distance of 200m. Simultaneously, 2 AEs hover at a horizontal distance of 100 meters from the BS. Besides the communication data streams transmitted to each user, 4 dedicated sensing beams are employed to sense and to interfere with the AEs. The number of BS antennas is set as $M_t=M_r=64$. The transmit power constraint is set as $P=30$ dBm, and the noise power is set as $\sigma_c^2=\sigma_e^2=\sigma_s^2=-80$ dBm. The small-scale fading is set as ${\left| {{\alpha _{n,i}}} \right|^2} = -10$ dB. We set the maximum allowable AEs' SINR and worst-case sensing SCNR as $\gamma_e = -10$ dB and $\gamma_r = 15$ dB, respectively. Besides, the RCS of each AE is set as $\zeta_t = 0$ dBsm.

The proposed beamforming designs for fully digital arrays and HAD arrays are labeled as `Proposed Digital Beamforming' and `Proposed HAD Beamforming'. Note that the secure beamforming matrices are designed based on the reconstructed AEs' CSI, and the sum secrecy rate is calculated by the perfect CSI to evaluate the practical performance. Moreover, we adopt the following schemes for comparison:

\begin{figure}[t]
	\centering
	\includegraphics[width=0.9\linewidth]{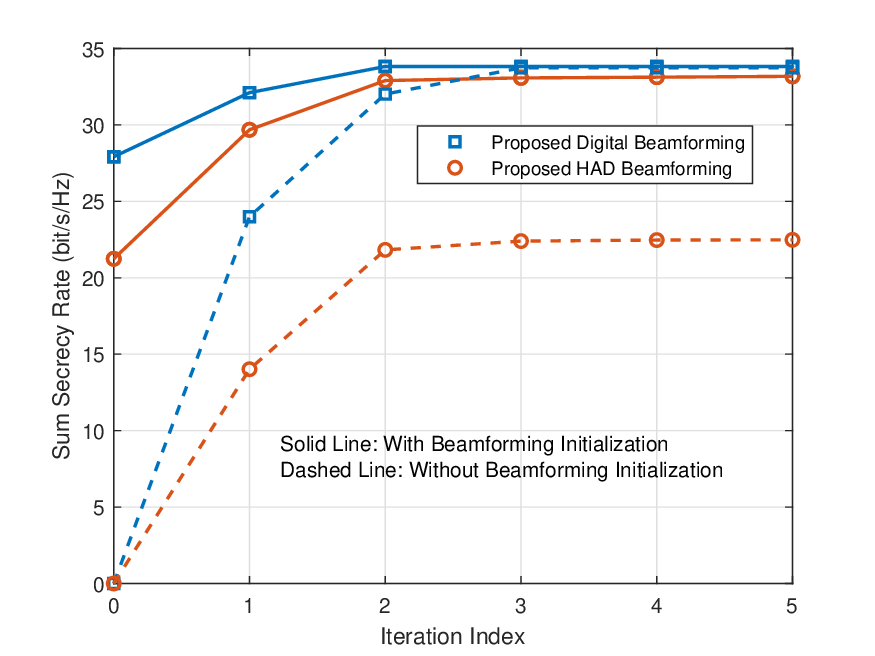}
	\caption{Convergence behavior of the proposed secure beamforming design algorithm.}
	\label{main_converge}
\end{figure}

\begin{figure}[]
	\centering
	\begin{subfigure}[]
		{\centering
			\includegraphics[width=0.48\linewidth]{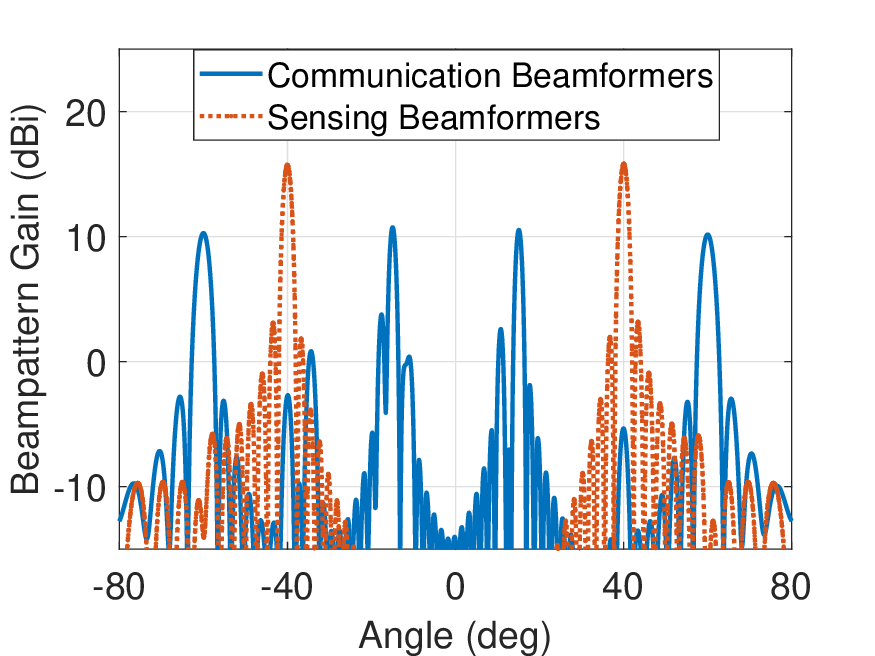}}
	\end{subfigure}
	\begin{subfigure}[]
		{\centering
			\includegraphics[width=0.48\linewidth]{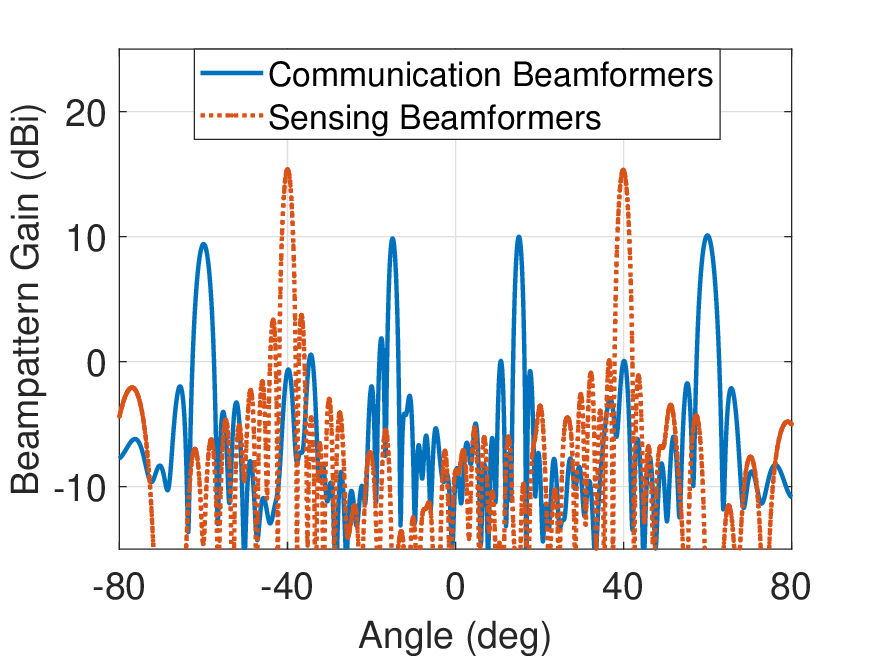}}
	\end{subfigure}
	\caption{Communication beampattern and sensing beampattern, $\gamma_r = 30$ dB. (a) Proposed digital beamforming. (b) Proposed HAD beamforming.}
	\label{Beampattern}
\end{figure}

\begin{itemize}
	\item Secure Beamforming with Perfect CSI \cite{8319269}: It is assumed that the perfect CSI of AEs is always available. The AN is employed to interfere with AEs. Compared with the proposed scheme, this scheme does not need to sense AEs, therefore performing as the upper bound.
	
	\item Communication-Only Beamforming: In this scheme, the BS transmits downlink communication signals to serve the communication users, while the echoes of the communication signals are utilized to sense the AEs and guide secure beamforming design. Note that this scheme does not employ extra sensing beams.
	
	\item Decomposed HAD Beamforming \cite{7397861}: The HAD beamforming matrices are decomposed based on the fully digital beamforming matrix, which is obtained from the proposed digital beamforming scheme.
\end{itemize}

In Fig. \ref{main_converge}, we first present the convergence behavior of the proposed algorithms. Both the `Proposed Digital Beamforming' and `Proposed HAD Beamforming' algorithms can converge within only a few iterations. As shown, with beamforming initialization (via \eqref{eq3.12a}, \eqref{eq3.12b}, and \eqref{eq4.16}), the initial sum secrecy rate is relatively high, thus enabling faster convergence of the algorithm. Moreover, a higher converged sum secrecy rate is achieved with effective initialization, since the obtained HAD beamforming matrix without initialization falls into a poor local optimum.


In Fig. \ref{Beampattern}, we compare the communication and sensing beampatterns of both fully digital arrays and HAD arrays. Here, the users and AEs are located at 
$\left[ { -60^\circ , - 15^\circ,15^\circ,60^\circ } \right]$ and $\left[ { - 40^\circ,40^\circ } \right]$, respectively. As can be seen, the communication and sensing beams are concentrated toward the directions of communication users and adversarial AEs, respectively. The sidelobes of the HAD beam are relatively higher than those of the fully digital beam due to limited degrees of freedom (DoFs), but it still maintains a high peak-to-sidelobe ratio.

\begin{figure}[t]
	\centering
	\includegraphics[width=0.9\linewidth]{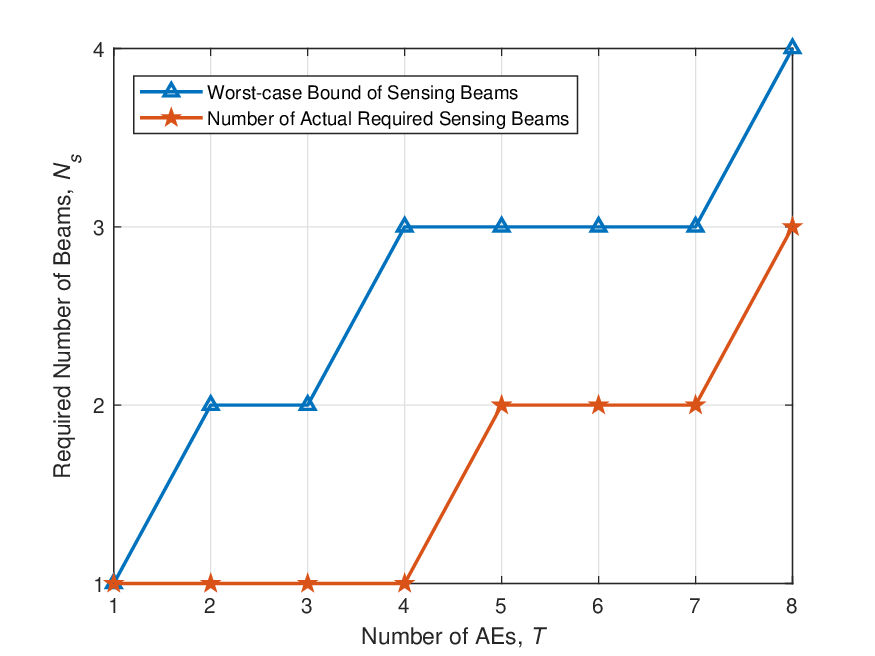}
	\caption{Required number of dedicated sensing beams.}
	\label{main_Ns_Theoretical_actual}
\end{figure}
In Fig. \ref{main_Ns_Theoretical_actual}, we investigate the required number of dedicated sensing beams for varying numbers of AEs. The worst-case bound of sensing beams is calculated by \eqref{eq3.11}, i.e., $N_s = \sqrt{2T+1}$, which scales with the square root of the number of AEs. Therefore, only a few sensing beams are sufficient to sense and interfere with the AEs. In the considered scenario, we validate this conclusion through numerical simulation, where the beamforming matrix is decomposed by the covariance matrix with rank reduction \cite{5233822}. It is observed that the actual required number of sensing beams is lower than the worst-case bound. This is because not all constraints in the optimization problem \eqref{eq2.14} are necessarily active at optimality. As shown in Fig. \ref{main_Ns_Theoretical_actual}, no more than four sensing beams are needed when the number of targets is smaller than eight. Accordingly, in the following evaluations, we consider four dedicated sensing beams.

\begin{figure}[t]
	\centering
	\includegraphics[width=0.9\linewidth]{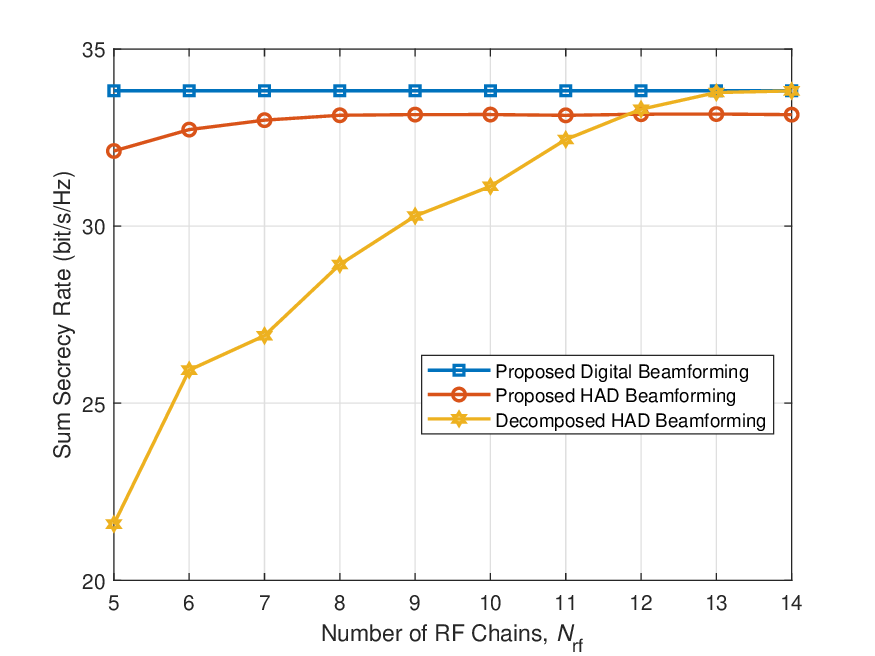}
	\caption{Sum secrecy rate versus the number of RF chains.}
	\label{main_Nrf_T2}
\end{figure}

In Fig. \ref{main_Nrf_T2}, we compare the sum secrecy rate under different numbers of RF chains with two AEs. It is observed that the `Proposed Digital Beamforming’ scheme achieves the best performance. Both `Proposed HAD Beamforming’ and `Decomposed HAD Beamforming’ exhibit performance improvement as the number of RF chains increases. The `Decomposed HAD Beamforming’ scheme exhibits excellent performance with sufficient RF chains. However, the performance degrades seriously with fewer RF chains, as its beamforming matrices cannot effectively approximate the fully digital beamforming matrix. Consequently, both the information leakage and the multi-user interference increase significantly. In contrast, the `Proposed HAD Beamforming’ maintains great performance even with limited RF chains due to the effective eavesdropping SINR constraints. Nevertheless, it is unfortunate that the sum secrecy rate does not increase significantly with more RF chains, as the local optima of the alternating algorithm become the primary performance bottleneck. Therefore, when the number of RF chains is relatively large, the `Decomposed HAD Beamforming' can be directly adopted to achieve a performance comparable to that of fully digital beamforming. In scenarios with fewer RF chains, the `Proposed HAD Beamforming' is more recommended to ensure reliable secure communication. In the following simulations, we focus on the scenarios with limited RF chains, where the number of RF chains is set to eight.

\begin{figure}[t]
	\centering
	\includegraphics[width=0.9\linewidth]{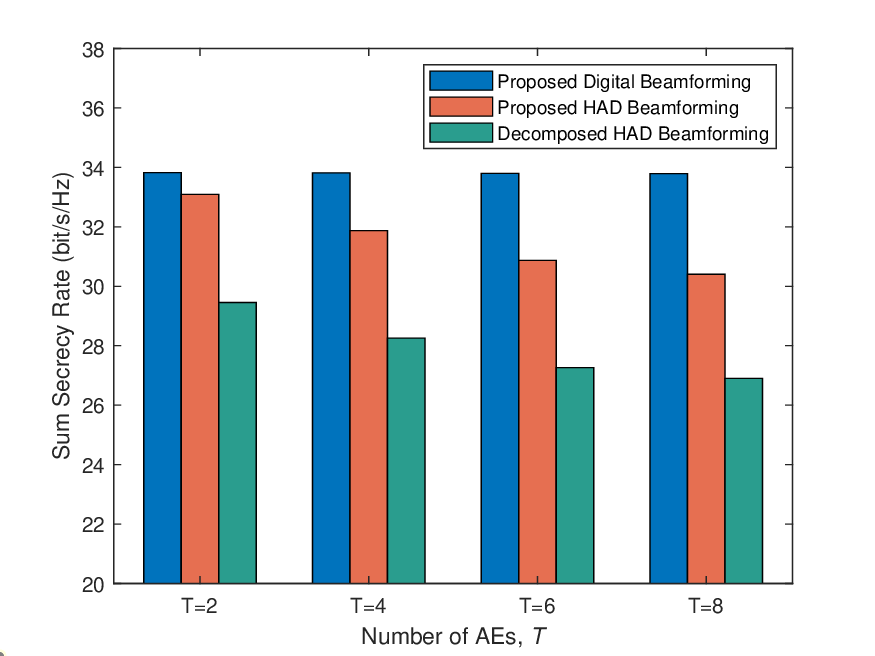}
	\caption{Sum secrecy rate versus the number of AEs.}
	\label{main_Nrf_T_bar}
\end{figure}

In Fig. \ref{main_Nrf_T_bar}, we compare the sum secrecy rate under different numbers of AEs. As shown, for the `Proposed Digital Beamforming' algorithm, the sum secrecy rate exhibits negligible degradation as the number of AEs increases, because the AEs' SINR constraint (\ref{eq2.14}a) in the optimization effectively prevents the information leakage. For the `Proposed HAD Beamforming' algorithm, the numerous constraints reduce the available DoFs, leading to slight performance degradation. \textcolor{black}{In contrast, the `Decomposed HAD Beamforming' scheme suffers a severe performance drop. Due to the lack of effective information leakage suppression, an increase in the number of eavesdroppers exacerbates information leakage.}



\begin{figure}[t]
	\centering
	\includegraphics[width=0.9\linewidth]{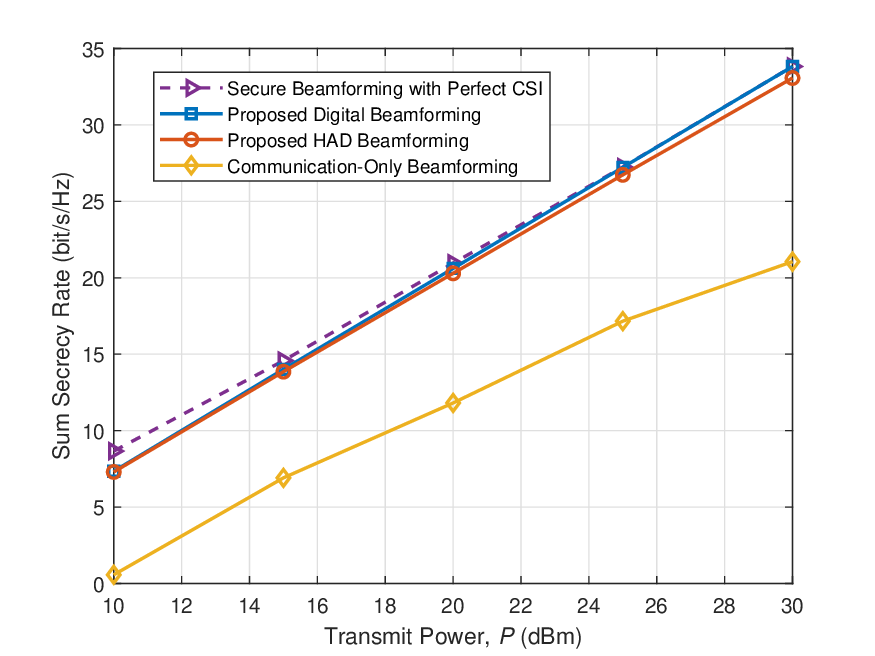}
	\caption{Sum secrecy rate achieved by different beamforming designs.}
	\label{main_Pk}
\end{figure}
\begin{figure}[t]
	\centering
	\includegraphics[width=0.9\linewidth]{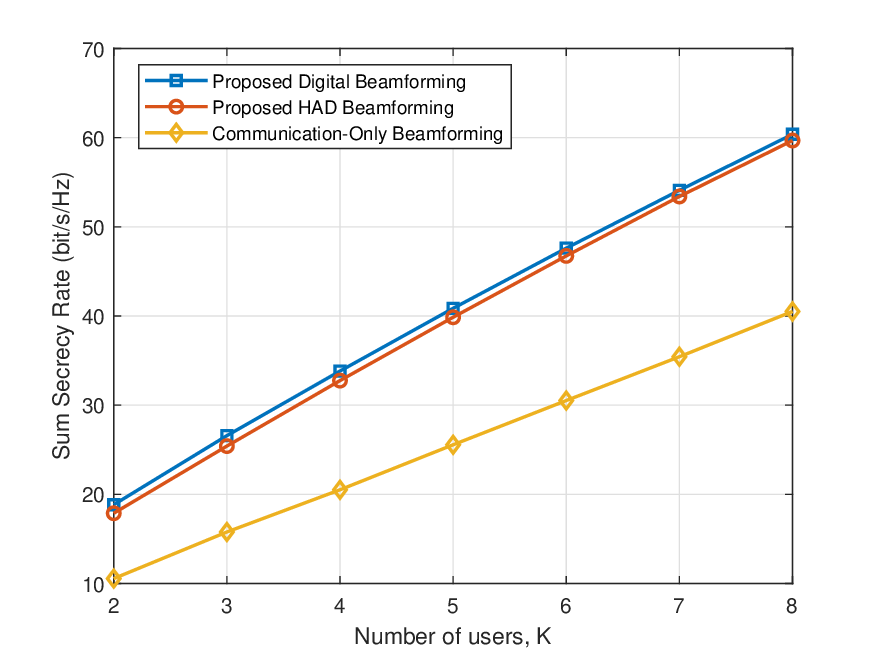}
	\caption{Sum secrecy rate versus the number of users.}
	\label{main_K}
\end{figure}

In Fig. \ref{main_Pk}, we compare the sum secrecy rate under different beamforming design algorithms. As can be seen, the `Secure Beamforming with Perfect CSI' achieves the best performance, since more resources are devoted to achieving higher secrecy rates, which serves as a performance upper bound. The `Proposed Digital Beamforming' algorithm performs slightly worse than the upper bound in the low transmit power region and approaches it as the power budget increases. Due to sensing SCNR constraints, sufficient power must be radiated toward the direction of the AEs, which degrades communication performance in the low transmit power region. While in the high transmit power region, a large amount of energy is used as AN to interfere with AE. Therefore, the sensing function no longer consumes additional energy. The performance of `Proposed HAD Beamforming'  is weaker than that of `Proposed Digital Beamforming', but it remains reasonably good. In contrast, the performance of the `Communication-Only Beamforming' scheme is significantly inferior to that of other algorithms. Since the information-carrying signal must align with AEs for sensing, this leads to inherent information leakage. Furthermore, in Fig. \ref{main_K}, it is shown that the sum secrecy rate increases approximately linearly with the number of users, which implies that the multi-user interference is well suppressed by the proposed beamforming design.

\begin{figure}[t]
	\centering
	\includegraphics[width=0.9\linewidth]{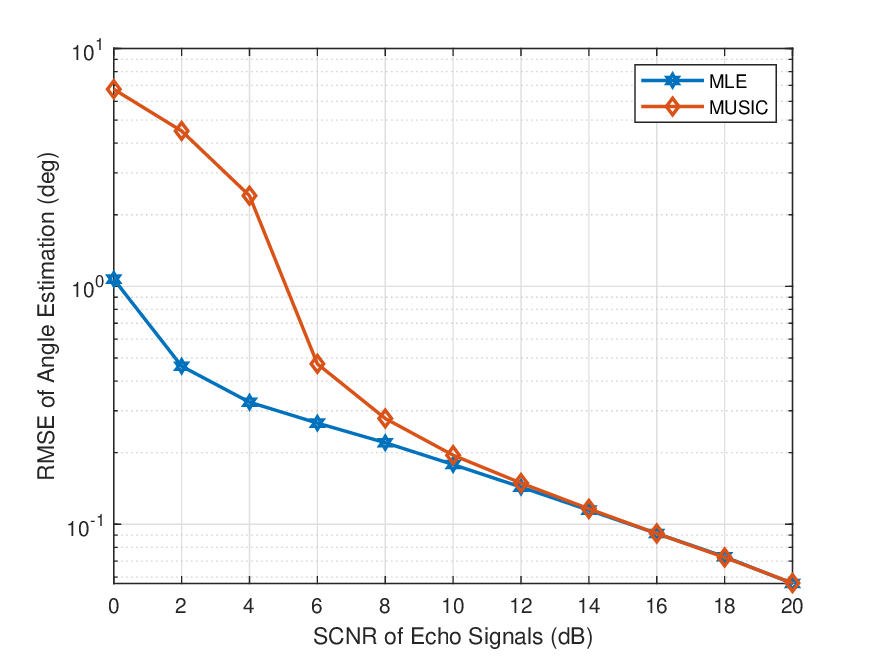}
	\caption{AEs' angle estimation RMSE under different sensing SCNR.}
	\label{main_DOA}
\end{figure}

\begin{figure}[t]
	\centering
	\includegraphics[width=0.9\linewidth]{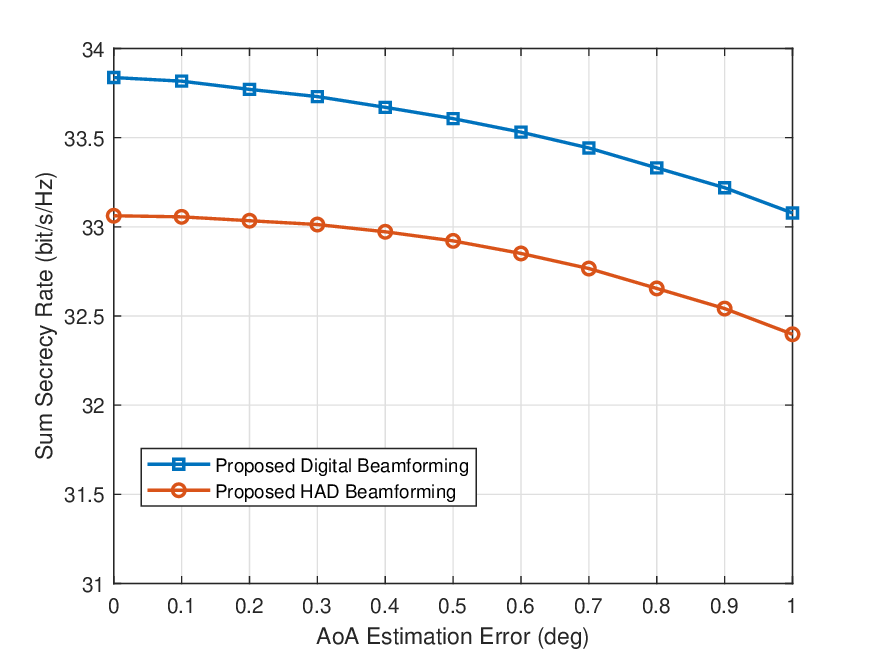}
	\caption{Sum secrecy rate versus the AEs' angle estimation RMSE.}
	\label{main_error}
\end{figure}

\textcolor{black}{In Fig. \ref{main_DOA}, we present the root mean squared error (RMSE) of the AEs' angle estimation under different sensing SCNR. As expected, the angle estimation RMSE with both MLE and MUSIC algorithms decreases as the SCNR increases, which proves that using SCNR as the sensing performance metric effectively improves the AEs estimation performance.}

\textcolor{black}{In Fig. \ref{main_error}, we further evaluate the impact of the AEs' angle estimation RMSE on secure communication. It is observed that the sum secrecy rate decreases with increasing angle estimation RMSE. This is because the dedicated sensing signals are misaligned with the AEs due to angles estimation errors, thereby reducing interference to the AEs and increasing information leakage. However, with the SCNR constraint set in this paper, i.e., $\gamma_r=15$ dB, the angle estimation RMSE is approximately reduced to 0.1° as shown in Fig. \ref{main_DOA}, and the resulting loss in sum secrecy rate is negligible.}



\begin{figure}[t]
	\centering
	\includegraphics[width=0.9\linewidth]{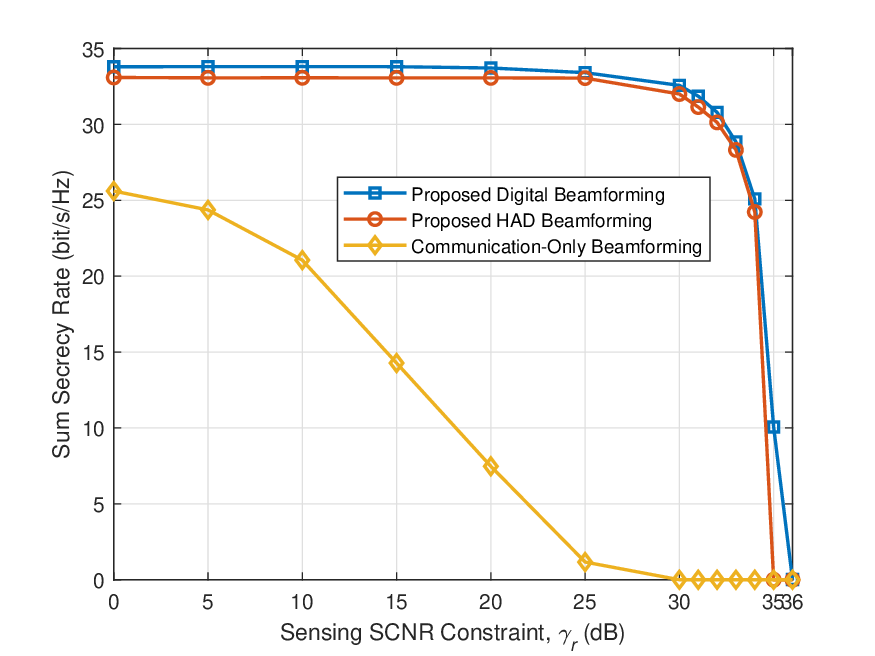}
	\caption{Trade-off between sum secrecy rate and sensing SCNR threshold.}
	\label{main_eta}
\end{figure}

At last, Fig. \ref{main_eta} shows the trade-off between the communication performance and sensing performance. 
As the sensing SCNR increases, for the `Communication-Only Beamforming' scheme, the received information-carrying signal power decreases at users, while it increases at the AEs, thus increasing the information leakage risk and degrading the communication performance. In contrast, for the proposed algorithm, the sum secrecy rate hardly decreases with the increase of sensing SCNR threshold when $\gamma_r \le 20$ dB. This is because the dedicated sensing signals need to be sufficiently strong to interfere with the AEs, even under loose sensing SCNR constraints. \textcolor{black}{This also ensures reliable sensing performance for AEs, which prevents the degradation of secure communication performance resulting from sensing errors of the AEs.} The sum secrecy rate only degrades when the sensing SCNR threshold becomes very high.

\section{Conclusion}
\textcolor{black}{This paper studied secure ISAC downlink transmission with joint design of confidential beams and a small number of dedicated sensing beams. For fully digital arrays, we proposed an FP-based alternating method with SCA, SDR, and closed-form auxiliary updates, established a worst-case rank bound that upper-bounds the minimum sensing-beam count. We then extended the design to HAD arrays via a penalty-based Riemannian manifold procedure. Simulation results indicate that only a few sensing beams are generally sufficient to sense/jam AEs while maintaining high secrecy rates; the proposed schemes outperform communication-only and decomposed-HAD baselines. The proposed HAD variant closely approaches digital performance, with moderate gaps due to RF-chain limitations. Varying the sensing-SCNR threshold reveals the communication–sensing tradeoff: dedicated beams are first allocated to satisfy sensing requirements, while significant power reallocation from communication occurs only at high thresholds. Overall, the results offer practical guidance on sizing sensing beams and designing a secure ISAC strategy under realistic hardware constraints.}


\appendices
\section{Proof of Theorem 1}
For any optimal covariance matrix of sensing beamforming matrix to problem \eqref{eq3.8} (or the original problem \eqref{eq2.14}), denoted as ${\bf R}_v$, we can always construct an alternative matrix ${\bf R}_v^\star$ that satisfies all the constraints and remains an optimal objective value. The feasibility of Theorem 1 is checked by solving the following problem:
	\begin{align}
		\label{appendix.1}
		\min _{{\bf{R}}_v^\star} ~~& {\rm Tr}\left( {{{{\bf{\tilde H}}}^H}{{\bf{R}}_v^\star}{\bf{\tilde H}}} \right) \\
		\mathrm{s.t.}\quad &{\bf{h}}_{e,t}^H{{\bf{R}}_v^\star}{{\bf{h}}_{e,t}} \ge {\bf{h}}_{e,t}^H{{\bf{R}}_v}{{\bf{h}}_{e,t}}, \forall t, \tag{\ref{appendix.1}a}\\
		& {\rm Tr}\left( {\left( {{{\bf{A}}_{{\bf{H}},t}} - {\gamma _r}{{\bf{A}}_{{\bf{H}},I}}} \right){{\bf{R}}_v^\star}} \right) \notag \\
		& \qquad \qquad \ge {\rm Tr}\left( {\left( {{{\bf{A}}_{{\bf{H}},t}} - {\gamma _r}{{\bf{A}}_{{\bf{H}},I}}} \right){{\bf{R}}_v}} \right), \forall t, \tag{\ref{appendix.1}b}\\ 
		& {\rm Tr}\left( {\bf R}_v^\star \right) \le {\rm Tr}\left( {\bf R}_v \right), \tag{\ref{appendix.1}c} \\
		& {\bf R}_v^\star \succeq 0, \tag{\ref{appendix.1}d} 
	\end{align}
	where the objective function of (\ref{appendix.1}) ensures that the interference caused by the dedicated sensing signal on users does not increase, thus maintaining the optimum objective value of \eqref{eq3.8}/\eqref{eq2.14}. Constraints (\ref{appendix.1}a)-(\ref{appendix.1}c) ensure that the optimized covariance of sensing beamforming matrix satisfy the AEs' SINR constraints, the sensing SCNR constraints, and the power budget constraint, corresponding to \eqref{eq3.4}/(\ref{eq2.14}a), \eqref{eq3.7}/(\ref{eq2.14}b), and (\ref{eq3.8}a)/(\ref{eq2.14}c), respectively. This problem is written in a more general form
	\begin{align}
		\label{appendix.2}
		\min _{{\bf{R}}_v^\star} ~~& {\rm Tr}\left({\bf A} {{\bf{R}}_v^\star}\right) \\
		\mathrm{s.t.}\quad &{\rm Tr}\left({\bf B}_m {{\bf{R}}_v^\star}\right) \trianglelefteq_m b_m, m = 1,\cdots,2T+1, \tag{\ref{appendix.1}a}  \\
		& {\bf R}_v^\star \succeq 0, \tag{\ref{appendix.2}b} 
	\end{align}
	where $\bf A$ and ${\bf B}_m, m = 1,\cdots,2T+1$, are Hermitian matrices, $\trianglelefteq_m \in \left\{ { \le , = , \ge } \right\}$. It is observed that this problem has $2T+1$ linear equations and a semipositive definite constraint, which determines the minimum rank of the optimal solution. Based on \cite{5233822,pataki1998rank}, the problem \eqref{appendix.2} always has an optimal solution obtained by the rank reduction method satisfying 
	\begin{align}
		\label{appendix.3}
		{\rm{rank}}\left( {{\bf{R}}_v^ \star } \right) \le \sqrt{2T+1}.
	\end{align}
	In practice, the rank of the $\mathbf{R}_v^\star$ may be much smaller than $\sqrt{2T+1}$, as not all constraints are necessarily active at optimality \cite{yao2025optimal}. Through decomposing ${\bf{R}}_v^ \star  = {{\bf{U}}_R}{{\bf{\Sigma }}_R}{\bf{U}}_R^H$, we can obtain the sensing beamforming matrix as ${\bf{V}} = {{\bf{U}}_R}{\bf{\Sigma }}_R^{1/2}{{\bf{I}}_{{M_t} \times {N_s}}}$, where ${{\bf{\Sigma }}_R} = {\rm{diag}}\left( {{\lambda _1},{\lambda _2}, \cdots,{\lambda _{{M_t}}}} \right)$ with diagonal elements being the eigenvalue arranged in descending order, ${{\bf{U}}_R}$ is the corresponding eigenvector matrix, and $N_s \ge {\rm{rank}}\left( {{\bf{R}}_v^ \star } \right)$ denotes the number of the sensing beams.

\bibliographystyle{IEEEtran}%
\bibliography{Secure_beamforming}

\begin{thebibliography}{10}
\providecommand{\url}[1]{#1}
\csname url@samestyle\endcsname
\providecommand{\newblock}{\relax}
\providecommand{\bibinfo}[2]{#2}
\providecommand{\BIBentrySTDinterwordspacing}{\spaceskip=0pt\relax}
\providecommand{\BIBentryALTinterwordstretchfactor}{4}
\providecommand{\BIBentryALTinterwordspacing}{\spaceskip=\fontdimen2\font plus
\BIBentryALTinterwordstretchfactor\fontdimen3\font minus
  \fontdimen4\font\relax}
\providecommand{\BIBforeignlanguage}[2]{{%
\expandafter\ifx\csname l@#1\endcsname\relax
\typeout{** WARNING: IEEEtran.bst: No hyphenation pattern has been}%
\typeout{** loaded for the language `#1'. Using the pattern for}%
\typeout{** the default language instead.}%
\else
\language=\csname l@#1\endcsname
\fi
#2}}
\providecommand{\BIBdecl}{\relax}
\BIBdecl

\bibitem{11006083}
M.~Ahmed, A.~A. Soofi, S.~Raza, Y.~Li, F.~Khan, W.~U. Khan, M.~Asif, and
  Z.~Han, ``A comprehensive survey on {RIS}-enhanced physical layer security in
  {UAV}-assisted networks,'' \emph{IEEE Internet Things J.}, pp. 1--1, Aug.
  2025.

\bibitem{10054167}
P.~Liu, Z.~Fei, X.~Wang, J.~A. Zhang, Z.~Zheng, and Q.~Zhang, ``Securing
  multi-user uplink communications against mobile aerial eavesdropper via
  sensing,'' \emph{IEEE Trans. Veh. Technol.}, vol.~72, no.~7, pp. 9608--9613,
  Jul. 2023.

\bibitem{10375133}
H.~Jia, X.~Li, and L.~Ma, ``Physical layer security optimization with
  {Cramér–Rao} bound metric in {ISAC} systems under sensing-specific
  imperfect {CSI} model,'' \emph{IEEE Trans. Veh. Technol.}, vol.~73, no.~5,
  pp. 6980--6992, May 2024.

\bibitem{10086626}
H.~Hua, J.~Xu, and T.~X. Han, ``Optimal transmit beamforming for integrated
  sensing and communication,'' \emph{IEEE Transactions on Vehicular
  Technology}, vol.~72, no.~8, pp. 10\,588--10\,603, Aug. 2023.

\bibitem{9737357}
F.~Liu, Y.~Cui, C.~Masouros, J.~Xu, T.~X. Han, Y.~C. Eldar, and S.~Buzzi,
  ``Integrated sensing and communications: Toward dual-functional wireless
  networks for {6G} and beyond,'' \emph{IEEE J. Sel. Areas Commun.}, vol.~40,
  no.~6, pp. 1728--1767, Jun. 2022.

\bibitem{7888145}
N.~González-Prelcic, R.~Méndez-Rial, and R.~W. Heath, ``Radar aided beam
  alignment in {mmWave} {V2I} communications supporting antenna diversity,'' in
  \emph{Inf. Theory Appl. Workshop (ITA)}, Mar. 2016, pp. 1--7.

\bibitem{8835615}
A.~Ali, N.~González-Prelcic, and A.~Ghosh, ``Millimeter wave {V2I}
  beam-training using base-station mounted radar,'' in \emph{Proc. IEEE Radar
  Conf. (RadarConf)}, Apr. 2019, pp. 1--5.

\bibitem{9171304}
F.~Liu, W.~Yuan, C.~Masouros, and J.~Yuan, ``Radar-assisted predictive
  beamforming for vehicular links: Communication served by sensing,''
  \emph{IEEE Trans. Wireless Commun.}, vol.~19, no.~11, pp. 7704--7719, Nov.
  2020.

\bibitem{10304580}
F.~Xia, Z.~Fei, J.~Huang, X.~Wang, R.~Wang, W.~Yuan, and D.~W.~K. Ng,
  ``Sensing-enabled predictive beamforming design for {RIS}-assisted {V2I}
  systems: A deep learning approach,'' \emph{IEEE Trans. Wireless Commun.},
  vol.~23, no.~6, pp. 5571--5586, Jun. 2024.

\bibitem{9593096}
N.~Su, Z.~Wei, and C.~Masouros, ``Secure dual-functional radar-communication
  system via exploiting known interference in the presence of clutter,'' in
  \emph{Proc. IEEE Workshop Signal Process. Adv. Wireless Commun. (SPAWC)},
  Sept. 2021, pp. 451--455.

\bibitem{6848758}
Z.~Chu, K.~Cumanan, Z.~Ding, M.~Johnston, and S.~Y. Le~Goff, ``Secrecy rate
  optimizations for a {MIMO} secrecy channel with a cooperative jammer,''
  \emph{IEEE Trans. Veh. Technol.}, vol.~64, no.~5, pp. 1833--1847, May 2015.

\bibitem{9838753}
Z.~Ren, L.~Qiu, and J.~Xu, ``Optimal transmit beamforming for secrecy
  integrated sensing and communication,'' in \emph{Proc. IEEE Int. Conf.
  Commun. (ICC)}, May 2022, pp. 5555--5560.

\bibitem{10153696}
Z.~Ren, L.~Qiu, J.~Xu, and D.~W.~K. Ng, ``Robust transmit beamforming for
  secure integrated sensing and communication,'' \emph{IEEE Trans. Commun.},
  vol.~71, no.~9, pp. 5549--5564, Sept. 2023.

\bibitem{10605793}
Z.~Ren, J.~Xu, L.~Qiu, and D.~Wing Kwan~Ng, ``Secure cell-free integrated
  sensing and communication in the presence of information and sensing
  eavesdroppers,'' \emph{IEEE Journal on Selected Areas in Communications},
  vol.~42, no.~11, pp. 3217--3231, 2024.

\bibitem{10227884}
N.~Su, F.~Liu, and C.~Masouros, ``Sensing-assisted eavesdropper estimation: An
  {ISAC} breakthrough in physical layer security,'' \emph{IEEE Trans. Wireless
  Commun.}, vol.~23, no.~4, pp. 3162--3174, Apr. 2024.

\bibitem{10639496}
K.~Hou and S.~Zhang, ``Optimal beamforming for secure integrated sensing and
  communication exploiting target location distribution,'' \emph{IEEE J. Sel.
  Areas Commun.}, vol.~42, no.~11, pp. 3125--3139, Nov. 2024.

\bibitem{10050406}
Z.~Wang, X.~Mu, and Y.~Liu, ``{STARS} enabled integrated sensing and
  communications,'' \emph{IEEE Trans. Wireless Commun.}, vol.~22, no.~10, pp.
  6750--6765, Oct. 2023.

\bibitem{10364735}
R.~Liu, M.~Li, Q.~Liu, and A.~Lee~Swindlehurst, ``{SNR/CRB}-constrained joint
  beamforming and reflection designs for {RIS-ISAC} systems,'' \emph{IEEE
  Trans. Wireless Commun.}, vol.~23, no.~7, pp. 7456--7470, Jul. 2024.

\bibitem{6966076}
E.~Zhang and C.~Huang, ``On achieving optimal rate of digital precoder by
  {RF}-baseband codesign for {MIMO} systems,'' in \emph{Proc. IEEE Veh.
  Technol. Conf. (VTC)}, Sept. 2014, pp. 1--5.

\bibitem{10619398}
C.~Xu and S.~Zhang, ``Integrated sensing and communication exploiting prior
  information: How many sensing beams are needed?'' in \emph{Proc. IEEE Int.
  Symp. Inf. Theor. (ISIT)}, Jul. 2024, pp. 2802--2807.

\bibitem{chen2020composite}
L.~Chen, F.~Liu, J.~Liu, and C.~Masouros, ``Composite signalling for {DFRC}:
  Dedicated probing signal or not?'' \emph{arXiv preprint arXiv:2009.03528},
  Sept. 2020.

\bibitem{10942665}
K.~M. Attiah and W.~Yu, ``Bounds on the minimum number of beamformers for
  integrated sensing and communications,'' in \emph{Conf. Rec. Asilomar Conf.
  Signals Syst. Comput. (ACSSC)}, Apr. 2024, pp. 1114--1118.

\bibitem{yao2025optimal}
J.~Yao and S.~Zhang, ``Optimal beamforming for multi-target multi-user {ISAC}
  exploiting prior information: How many sensing beams are needed?''
  \emph{arXiv preprint arXiv:2503.03560}, Mar. 2025.

\bibitem{9857564}
P.~Liu, Z.~Fei, X.~Wang, B.~Li, Y.~Huang, and Z.~Zhang, ``Outage constrained
  robust secure beamforming in integrated sensing and communication systems,''
  \emph{IEEE Wireless Commun. Lett.}, vol.~11, no.~11, pp. 2260--2264, Nov.
  2022.

\bibitem{richards2005fundamentals}
M.~A. Richards \emph{et~al.}, \emph{Fundamentals of radar signal
  processing}.\hskip 1em plus 0.5em minus 0.4em\relax Mcgraw-hill New York,
  2005, vol.~1.

\bibitem{van2002optimum}
H.~L. Van~Trees, \emph{Optimum array processing: {Part IV} of detection,
  estimation, and modulation theory}.\hskip 1em plus 0.5em minus 0.4em\relax
  John Wiley \& Sons, 2002.

\bibitem{7463025}
Y.~Sun, D.~W.~K. Ng, J.~Zhu, and R.~Schober, ``Multi-objective optimization for
  robust power efficient and secure full-duplex wireless communication
  systems,'' \emph{IEEE Trans. Wireless Commun.}, vol.~15, no.~8, pp.
  5511--5526, Aug. 2016.

\bibitem{9652071}
F.~Liu, Y.-F. Liu, A.~Li, C.~Masouros, and Y.~C. Eldar, ``Cramér-rao bound
  optimization for joint radar-communication beamforming,'' \emph{IEEE Trans.
  Signal Process.}, vol.~70, pp. 240--253, Dec. 2022.

\bibitem{11105450}
P.~Liu, S.~Xu, S.~Tang, X.~Wang, F.~Xia, W.~Yuan, and Z.~Fei, ``Sensing
  assisted secure communications: A rate-splitting approach,'' \emph{IEEE
  Internet Things J.}, pp. 1--1, 2025.

\bibitem{8314727}
K.~Shen and W.~Yu, ``Fractional programming for communication systems—{Part
  I}: Power control and beamforming,'' \emph{IEEE Trans. Signal Process.},
  vol.~66, no.~10, pp. 2616--2630, May 2018.

\bibitem{razaviyayn2013unified}
M.~Razaviyayn, M.~Hong, and Z.-Q. Luo, ``A unified convergence analysis of
  block successive minimization methods for nonsmooth optimization,''
  \emph{SIAM J. Optim.}, vol.~23, no.~2, pp. 1126--1153, Sept. 2012.

\bibitem{5233822}
Y.~Huang and D.~P. Palomar, ``Rank-constrained separable semidefinite
  programming with applications to optimal beamforming,'' \emph{IEEE Trans.
  Signal Process.}, vol.~58, no.~2, pp. 664--678, Feb. 2010.

\bibitem{7397861}
X.~Yu, J.-C. Shen, J.~Zhang, and K.~B. Letaief, ``Alternating minimization
  algorithms for hybrid precoding in millimeter wave {MIMO} systems,''
  \emph{IEEE J. Sel. Top. Signal Process.}, vol.~10, no.~3, pp. 485--500, Apr.
  2016.

\bibitem{5447068}
Z.-q. Luo, W.-k. Ma, A.~M.-c. So, Y.~Ye, and S.~Zhang, ``Semidefinite
  relaxation of quadratic optimization problems,'' \emph{IEEE Signal Process.
  Mag.}, vol.~27, no.~3, pp. 20--34, May 2010.

\bibitem{Absil2009Optimization}
P.-A. Absil, R.~Mahony, and R.~Sepulchre, ``Optimization algorithms on matrix
  manifolds,'' in \emph{Optimization Algorithms on Matrix Manifolds}.\hskip 1em
  plus 0.5em minus 0.4em\relax Princeton University Press, Apr. 2009.

\bibitem{9868348}
X.~Wang, Z.~Fei, J.~A. Zhang, and J.~Xu, ``Partially-connected hybrid
  beamforming design for integrated sensing and communication systems,''
  \emph{IEEE Trans. Commun.}, vol.~70, no.~10, pp. 6648--6660, Oct. 2022.

\bibitem{8319269}
W.~Kim, S.~Ha, J.~Koh, and J.~Kang, ``Artificial noise-aided secure beamforming
  for multigroup multicast,'' in \emph{Proc. IEEE Annu. Consum. Commun. Netw.
  Conf. (CCNC)}, Jan. 2018, pp. 1--4.

\bibitem{pataki1998rank}
G.~Pataki, ``On the rank of extreme matrices in semidefinite programs and the
  multiplicity of optimal eigenvalues,'' \emph{Math. Oper. Res.}, vol.~23,
  no.~2, pp. 339--358, May 1998.

\end{thebibliography}

\end{document}